\documentclass[twocolumn]{aastex62}
\usepackage{graphicx}
\usepackage{natbib}
\usepackage{placeins}
\usepackage{amsmath}
\usepackage{tgcursor}
\usepackage{booktabs}

\setcitestyle{notesep={ }}

\begin{document}

\title{A survey of C$_2$H, HCN, and C$^{18}$O in protoplanetary disks}

\author{Jennifer B. Bergner}
\affiliation{Harvard University Department of Chemistry and Chemical Biology, Cambridge, MA 02138, USA}

\author{Karin I. \"Oberg}
\affiliation{Harvard-Smithsonian Center for Astrophysics, Cambridge, MA 02138, USA}

\author{Edwin A. Bergin}
\affiliation{ University of MIchigan Department of Astronomy, Ann Arbor, MI 48109, USA}

\author{Ryan A. Loomis}
\affiliation{National Radio Astronomical Observatory, Charlottesville, VA 22903, USA}

\author{Jamila Pegues}
\affiliation{Harvard-Smithsonian Center for Astrophysics, Cambridge, MA 02138, USA}

\author{Chunhua Qi}
\affiliation{Harvard-Smithsonian Center for Astrophysics, Cambridge, MA 02138, USA}

\begin{abstract}
\noindent Molecular lines observed towards protoplanetary disks carry information about physical and chemical processes associated with planet formation.  We present ALMA Band 6 observations of C$_2$H, HCN, and C$^{18}$O in a sample of 14 disks spanning a range of ages, stellar luminosities, and stellar masses.  Using C$_2$H and HCN hyperfine structure fitting and HCN/H$^{13}$CN isotopologue analysis, we extract optical depth, excitation temperature, and column density radial profiles for a subset of disks.  C$_2$H is marginally optically thick ($\tau$ $\sim$1--5) and HCN is quite optically thick ($\tau$ $\sim$ 5--10) in the inner 200 AU.  The extracted temperatures of both molecules are low (10--30K), indicative of either sub-thermal emission from the warm disk atmosphere or substantial beam dilution due to chemical substructure.  We explore the origins of C$_2$H morphological diversity in our sample using a series of toy disk models, and find that disk-dependent overlap between regions with high UV fluxes and high atomic carbon abundances can explain a wide range of C$_2$H emission features (e.g. compact vs. extended and ringed vs. ringless emission).  We explore the chemical relationship between C$_2$H, HCN, and C$^{18}$O and find a positive correlation between C$_2$H and HCN fluxes, but no relationship between C$_2$H or HCN with C$^{18}$O fluxes.  We also see no evidence that C$_2$H and HCN are enhanced with disk age.  C$_2$H and HCN seem to share a common driver, however more work remains to elucidate the chemical relationship between these molecules and the underlying evolution of C, N, and O chemistries in disks.
\end{abstract}


\section{Introduction}
The compositions of young planets are set by the content of gas and solids in the protoplanetary disks from which they form.  Across a disk, volatile elements (C, N, O) are partitioned among different molecular carriers, which are in turn partitioned between the gas and ice phases.  As a result, the ratio of volatile elements in the gas and ice phase varies spatially and temporally in the disk. This chemical stratification is proposed to contribute to the range of planet compositions observed in our Solar System and extra-Solar systems \citep[e.g.][]{Madhusudhan2011, Oberg2011, Madhusudhan2012, Piso2016, Cridland2016, Oberg2016}.

In recent years, observations of small molecules in disks have provided important clues as to how this chemistry plays out.  Growing evidence indicates that CO is depleted in disks relative to interstellar abundances, even in warm disk layers above the CO freeze-out zone \citep{Dutrey2003, Chapillon2008, Miotello2017}.  HD observations have confirmed that this CO depletion is not simply due to a reduced gas mass, but is a true chemical effect \citep{Bergin2013, Favre2013, McClure2016, Schwarz2016}.  Moreover, gas-phase atomic carbon is also found to be depleted in disks \citep{Kama2016}, indicating a global rather than CO-specific gas-phase carbon depletion.  H$_2$O observations reveal a similar story, with very weak emission indicating heavy depletion of gas-phase water \citep{Bergin2010, Hogerheijde2011, Du2015, Du2017}.  Modeling of small molecule emission in IM Lup indicates that oxygen is more heavily depleted from the molecular layer than carbon, suggesting that at least in this young disk, oxygen is depleted first \citep{Cleeves2018}.

There is compelling evidence that the mechanism underlying volatile depletion involves grain growth and settling of ice-coated dust grains to the midplane \citep[e.g.][]{Bergin2010, Du2015, Du2017}.  In this framework, larger particles formed by efficient agglomeration of smaller icy grains settle to the midplane, carrying reservoirs of frozen volatile molecules.  Small particles may be cycled up from the midplane to the disk atmosphere, but larger particles will remain trapped in the midplane.  The transport of volatiles into the midplane is further aided by diffusive flows due to density gradients across the snow surface \citep{Meijerink2009, Xu2017}.  In the midplane, volatiles can be permanently sequestered by mechanisms such as grain growth \citep{Krijt2016, Krijt2018} or chemical transformation to form larger species \citep[e.g.][]{Walsh2014,Reboussin2015,Eistrup2016,Schwarz2018}.  Thus, oxygen and carbon can be depleted from disk atmospheres without altering the total gas or dust mass of the disk.

The depletion of oxygen and carbon from the gas phase is expected to lead to efficient formation of cyanide, hydrocarbon, and hydronitrogen molecules (CN-bearing, CH-bearing, and NH-bearing, respectively) in the disk atmosphere \citep{Du2015}.  This has been invoked to explain the bright emission of C$_2$H and C$_3$H$_2$ \citep{Bergin2016} as well as CH$_3$CN and HC$_3$N \citep{Bergner2018, Loomis2018} in disks.  To date, however, there are very few constraints on how CO, hydrocarbon, and cyanide chemistries are related within a single disk.  Recent work by \citet{Cleeves2018} provides some clues: using observations and modeling of C$_2$H and HCN in IM Lup, the authors find evidence for significant oxygen depletion, moderate carbon depletion, and no nitrogen depletion in the warm molecular layer.  This fits well with a picture of H$_2$O and CO sequestration in the midplane, though additional observations are required to test the relationship between these chemical families in a broader disk sample.

To this end, we present a survey of C$^{18}$O, C$_2$H, and HCN in 14 protoplanetary disks.  In addition to tracing the oxygen, carbon, and nitrogen chemistries, observations of these molecules may encode other interesting features of disk chemistry and physics.  For instance, in DM Tau and TW Hya, C$_2$H exhibits ringed emission external to the (sub-)mm dust continuum \citep{Kastner2015, Bergin2016}.  This morphology is thought to be sensitive to the strength of the UV field as well as oxygen and carbon depletion \citep{Bergin2016, Cleeves2018}, and therefore may have an important connection to the physical environment in the disk.  Additionally, small molecules have been proposed as possible tracers of planet formation, for instance via HCN sublimation induced by local heating of the disk by a planet \citep{Cleeves2015}.  Better observational constraints on the chemistry of small molecules are required to evaluate their utility as probes of planet formation.

In Section \ref{sec:obs} we describe the source sample, ALMA and SMA observations, and data reduction strategy.  Observational results, including molecule detections and emission morphologies, are presented in Section \ref{sec:obs_res}.  In Section \ref{sec:tau_cd} we extract C$_2$H and HCN optical depths and column densities using hyperfine structure fitting and isotopologue ratios.  In Section \ref{sec:diskmodels} we construct toy models of the local UV fluxes and atomic carbon abundances in disks to interpret the observed C$_2$H morphologies.  Lastly, in Section \ref{sec:disc} we discuss implications of the derived C$_2$H and HCN excitation temperatures and column density profiles and explore the chemical relationship between C$_2$H, HCN, and C$^{18}$O.

\section{Observations}
\label{sec:obs}
\subsection{Observational details}
Source details for the 14 protoplanetary disks in our sample are listed in Table \ref{tab:srcdat}.  The target disks were chosen to span a range of ages ($<$1 to $\sim$10 Myr), with the sample divided roughly in half between young ($<$5 Myr) and old ($>$5 Myr) disks, enabling us to test the evolutionary impact on the O, C, and N chemistries.  The sample also varies by over an order of magnitude in stellar mass and stellar luminosity, enabling an exploration of how disk physical properties impact the observed small molecule chemistry.

This project makes use of data from various ALMA and SMA programs, summarized in Table \ref{tab:obsdat}.  Full  details for the C$_2$H and HCN observations can be found in Appendix \ref{sec:app_obsdat}.  C$^{18}$O observational details are described in \citet{Huang2017} and J. Pegues (in preparation).  For all disks, four C$_2$H hyperfine transitions are covered: N = 3 -- 2, J = $\frac{7}{2}$ -- $\frac{5}{2}$, F = 4 -- 3 and F = 3 -- 2; and N = 3 -- 2, J = $\frac{5}{2}$ -- $\frac{3}{2}$, F = 3 -- 2 and F = 2 -- 1.  The HCN J = 3 -- 2 transition is also covered for all disks: CI Tau, DM Tau, DO Tau, LkCa 15, and MWC 480 were observed with the SMA and the remaining disks with ALMA.  H$^{13}$CN J = 3 -- 2 is included for AS 209, CI Tau, DM Tau, DO Tau, HD 163296, LkCa 15, MWC 480, and V4046 Sgr.  Lastly, C$^{18}$O J = 2 -- 1 was targeted towards all disks, together with CO J = 2 -- 1.  We present C$^{18}$O when it is detected, and otherwise CO.

\begin{deluxetable*}{lclllcc} 
	\tabletypesize{\footnotesize}
	\tablecaption{Source targets: disk \& star properties \label{tab:srcdat}}
	\tablecolumns{7} 
	\tablewidth{\textwidth} 
	\tablehead{
		\colhead{Source}                          &
		\colhead{Dist.$^a$}                       &
		\colhead{Age}                               & 
		\colhead{$M_\star$}                     & 
		\colhead{$L_\star$}                      & 
		\multicolumn{2}{c}{Keplerian mask parameters$^b$}\\
		\colhead{}                                     & 
		\colhead{(pc)}                               &
		\colhead{(Myr)}                            & 
		\colhead{($M_\odot$)}                 & 
		\colhead{($L_\odot$)}                 &
		\colhead{Inc. (deg)}                           & 
		\colhead{PA (deg)}                           
		  }
\startdata
AS 209 & 121.0 & 1.6$^{[1]}$ & 0.9$^{[1]}$ & 1.5$^{[1]}$ & 50 & 91 \\
CI Tau & 158.7 & 1.5 -- 2.5$^{[2, 3]}$ & 0.8$^{[2, 3]}$ & 0.9$^{[3]}$ & 50 & 14 \\
DM Tau & 145.1 & 3.5 -- 8$^{[2, 3]}$ & 0.5 -- 0.6$^{[3, 4]}$ & 0.2$^{[3]}$ & 45 & 64 \\
DO Tau & 139.4 & 0.8$^{[3]}$ & 0.5$^{[3]}$ & 1.4$^{[3]}$ & 30 & 5 \\
HD 143006 & 166.1 & 1$^{[5]}$ & 1$^{[6]}$ & 0.8$^{[6]}$ & 15 & 12 \\
HD 163296 & 101.5 & 7.6$^{[7]}$ & 1.8 $^{[7]}$ & 15.8 $^{[7]}$ & 49 & 48 \\
IM Lup & 158.4 & 1$^{[8]}$ & 1.0$^{[9]}$ & 1.7$^{[9]}$ & 50 & 142 \\
J1604-2130 & 150.1 & 5 -- 11$^{[10, 11]}$ & 0.9$^{[12]}$ & 0.6$^{[12]}$ & 5 & 86 \\
J1609-1908 & 137.6 & 5 -- 11$^{[10, 11]}$ & 0.8$^{[12]}$ & 0.4$^{[12]}$ & 55 & 68 \\
J1612-1859 & 139.1 & 5 -- 11$^{[10, 11]}$ & 0.5$^{[13]}$ & 0.3$^{[13]}$ & 60 & 35 \\
J1614-1906 & 143.9 & 5 -- 11$^{[10, 11]}$ & 1.0 $^{[12]}$& 0.5$^{[12]}$ & 55 & 7 \\
LkCa 15 & 158.9 & 3 -- 5$^{[2, 3, 4]}$ & 1.05$^{[3]}$; 0.97$^{[4]}$ & 0.8$^{[3]}$ & 50 & 61 \\
MWC 480 & 161.8 & 6.2$^{[7]}$ & 1.8$^{[7]}$ & 18.6$^{[7]}$ & 35 & 53 \\
V4046 Sgr & 72.4 & 4 -- 13$^{[14, 15]}$& 0.9, 0.7$^{[15]}$ & 0.49, 0.33$^{[15]}$ & 40 & 76 \\
\enddata
\tablenotetext{a}{From \textit{Gaia} Data Release 2 \citep{Gaia2018}}
\tablenotetext{b}{Derived from Keplerian mask fitting to CO emission (J. Pegues, in preparation)}
\tablenotetext{}{References: [1] \citet{Andrews2009}, [2] \citet{Guilloteau2014}, [3] \citet{Andrews2013},  [4] \citet{Simon2000},
[5] \citet{Dunkin1997}, [6] \citet{Natta2004}, [7] \citet{Vioque2018},  [8] \citet{Galli2015}, [9] \citet{Tazzari2017}, [10] \citet{Preibisch2002}, [11] \citet{Pecaut2012}, [12] \citet{Carpenter2014}, [13] \citep{Barenfeld2016}, [14] \citet{Rosenfeld2012}, 
[15] \citet{Quast2000}}
\end{deluxetable*}

\begin{deluxetable*}{llll} 
	\tabletypesize{\footnotesize}
	\tablecaption{Observation summary \label{tab:obsdat}}
	\tablecolumns{4} 
	\tablewidth{\textwidth} 
	\tablehead{
		\colhead{Telescope}               &
		\colhead{Project code}           &
		\colhead{Molecule targets}     &
		\colhead{Sources}  }
\startdata
ALMA & 2013.1.00226.S & H$^{13}$CN & V4046 Sgr \\
 & & C$^{18}$O & AS 209, HD 163296, IM Lup, LkCa 15, MWC 480, V4046 Sgr \\
ALMA & 2015.1.00964.S & C$_2$H, HCN & AS 209, IM Lup, HD 163296, HD 143006, J1604-2130, \\
 & & & J1609-1908, J1614-1906, J1612-1859 \\
 & & C$^{18}$O, CO & HD 143006, J1604-2130, J1609-1908, J1614-1906, J1612-1859 \\
 ALMA & 2015.1.00671.S & HCN, C$_2$H & V4046 Sgr  \\
ALMA & 2016.1.00627.S & C$_2$H, H$^{13}$CN  & CI Tau, DM Tau, DO Tau, LkCa 15, MWC 480\\
 &  & C$^{18}$O & CI Tau, DM Tau, DO Tau \\
SMA & 2017B-S036 & HCN & CI Tau, DO Tau, DM Tau, LkCa 15, MWC 480 
\enddata
\end{deluxetable*}

ALMA data were self-calbrated in CASA using the continuum emission from individual spectral windows.  One or  two rounds of phase self-calibration were performed depending on the continuum brightness.  Continuum imaging was performed using Briggs weighting with a robust factor of 0.5.  The self-calibration solutions were applied to the spectral line data followed by continuum subtraction in the uv plane.  SMA data were calibrated using the \texttt{MIR~} software package \footnote{http://www.cfa.harvard.edu/$\sim$cqi/mircook.html}, then exported to fits and converted into CASA measurement sets.  For both ALMA and SMA observations, calibrated and continuum-subtracted data were CLEANed to a 3$\sigma$ noise threshold in CASA 4.6, using Briggs weighting with a robust factor of 1.0.  CLEAN masks were drawn by hand for strong emission; for weak lines the 5$\sigma$ continuum contour was used as the CLEAN mask.  Channel maps were generally produced with a spectral resolution of 0.5 km/s.  To produce image cubes for hyperfine fitting a higher spectral resolution of 0.14 km/s was used.

\subsection{Data analysis}
\label{sec:datared}
\begin{figure*}
	\includegraphics[width=\linewidth]{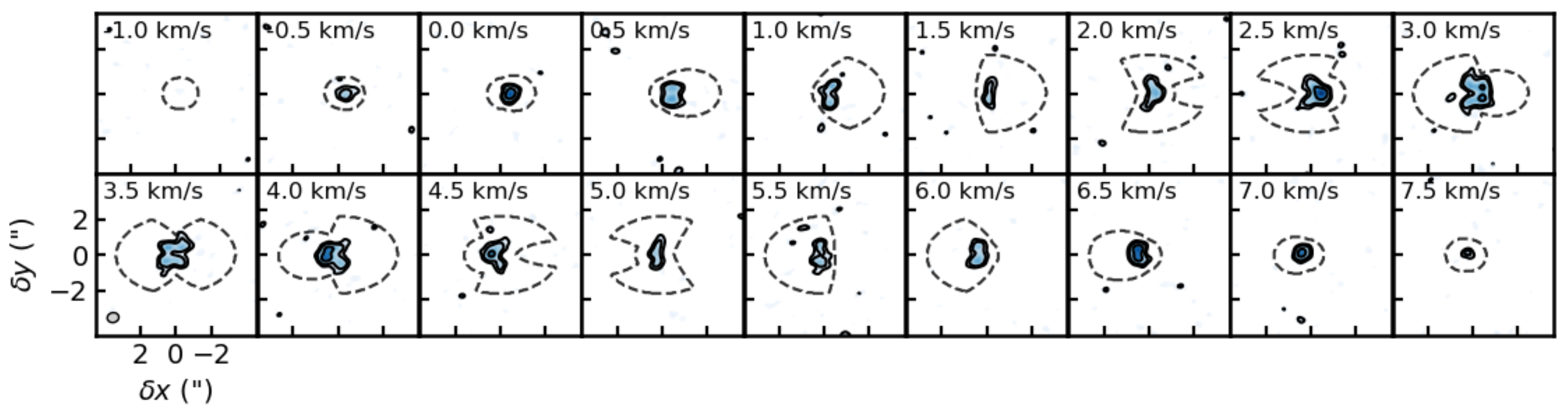}
	\caption{Channel map of the C$_2$H N = 3 -- 2, J = $\frac{7}{2}$ -- $\frac{5}{2}$, F = 4 -- 3 and 3 -- 2 transitions in AS 209 with an overlaid Keplerian mask.  Contours correspond to 3, 5, and 10 $\sigma$ emission.  The restoring beam is shown in the bottom left panel.  All additional channel maps are available as a figure set in the online journal.}
\label{fig:chan_as209_c2h}
\end{figure*}

We use Keplerian masking to extract moment zero maps, line spectra, and integrated flux densities for each target line \citep{Salinas2017}.  Our methodology is described in detail in \citet{Bergner2018} and J. Pegues (in preparation).  Briefly, masks were constructed using the disk and star physical parameters listed in Table \ref{tab:srcdat}.  We adopt an outer radius based on the extent of emission in the deprojected radial profiles of each line (Figure \ref{fig:radprof}).  Figure \ref{fig:chan_as209_c2h} shows an example channel map with the Keplerian mask overlaid. 

Line observation results for the HCN J = 3 -- 2 and C$_2$H N = 3 --2, J = $\frac{7}{2}$ -- $\frac{5}{2}$ transitions are listed in Table \ref{tab:fluxes_hcn_c2h}.  Additional results for H$^{13}$CN J = 3 -- 2, C$^{18}$O and CO J = 2 -- 1, and C$_2$H N = 3 --2, J = $\frac{5}{2}$ -- $\frac{3}{2}$ are listed in Appendix \ref{sec:app_obsdat}.  We estimate the integrated flux density uncertainties by bootstrapping: the same Keplerian mask used to extract line emission is applied to 1000 synthetic image cubes composed of randomly drawn emission-free channels neighboring the target line.  Integrated flux uncertainties are calculated from the standard deviation in integrated fluxes in these emission-free moment zero images, added in quadrature with a 10\% calibration uncertainty.  

\begin{deluxetable*}{lccccc} 
	\tabletypesize{\footnotesize}
	\tablecaption{HCN and C$_2$H Line Observations (Detections and Nondetection Upper Limits) \label{tab:fluxes_hcn_c2h}}
	\tablecolumns{6} 
	\tablewidth{\textwidth} 
	\tablehead{
		\colhead{Source}               &
		\colhead{Beam}                 &
		\colhead{Beam PA}           &
		\colhead{Channel rms$^a$}      &
		\colhead{Mom. Zero rms$^b$}  &
		\colhead{Int. Flux Density$^c$}  \\ 
		\colhead{}                           & 
		\colhead{('')}                        &
		\colhead{($^{\rm{o}}$)}          &
		\colhead{(mJy beam$^{-1}$)} &
		\colhead{(mJy beam$^{-1}$ km s$^{-1}$)} & 
		\colhead{(mJy km s$^{-1}$)}                      
		}
\startdata
\multicolumn{6}{c}{HCN J = 3--2} \\
\hline
AS 209 & 0.56 $\times$ 0.50 & 62.9 & 6.3 & 13.6 & 3707 $\pm$ 380\\
CI Tau$^d$ & 1.32 $\times$ 1.02 & 89.6 & 61.5 & 108.5 & 1271 $\pm$ 390\\
DM Tau$^d$ & 1.32 $\times$ 1.01 & 89.5 & 57.7 & 60.3 & 2966 $\pm$ 400\\
DO Tau$^d$ & 1.68 $\times$ 1.37 & -86.0 & 78.7 & 131.1 & $<$ 1252\\
HD 163296 & 0.53 $\times$ 0.47 & 84.3 & 2.3 & 6.1 & 9174 $\pm$ 920\\
IM Lup & 0.55 $\times$ 0.53 & -11.9 & 7.6 & 12.5 & 5654 $\pm$ 570\\
LkCa 15$^d$ & 2.92 $\times$ 2.58 & -70.4 & 99.6 & 177.0 & 3809 $\pm$ 450\\
MWC 480$^d$ & 2.88 $\times$ 2.61 & -67.1 & 90.3 & 230.1 & 1863 $\pm$ 390\\
HD 143006 & 0.55 $\times$ 0.44 & -85.5 & 5.4 & 9.7 & 2059 $\pm$ 210\\
J1604-2130 & 0.56 $\times$ 0.44 & -89.0 & 5.5 & 9.0 & 9709 $\pm$ 970\\
J1609-1908 & 0.54 $\times$ 0.44 & -86.8 & 5.4 & 10.5 & 1052 $\pm$ 110\\
J1612-1859 & 0.55 $\times$ 0.44 & -86.6 & 5.2 & 9.6 & $<$ 93\\
J1614-1906 & 0.55 $\times$ 0.44 & -86.3 & 5.3 & 9.9 & $<$ 97\\
V4046 Sgr & 0.77 $\times$ 0.57 & 76.1 & 5.3 & 11.7 & 10000 $\pm$ 1000\\
\hline
\multicolumn{6}{c}{C$_2$H N = 3--2, J = $\frac{7}{2}$ -- $\frac{5}{2}$, F = 4 --3 and 3--2} \\
\hline
AS 209 & 0.65 $\times$ 0.55 & 77.6 & 10.3 & 17.8 & 2709 $\pm$ 290\\
CI Tau & 0.59 $\times$ 0.49 & -22.1 & 3.0 & 5.2 & 1041 $\pm$ 110\\
DM Tau & 0.57 $\times$ 0.49 & -32.3 & 3.2 & 4.5 & 2013 $\pm$ 200\\
DO Tau & 0.63 $\times$ 0.49 & -21.8 & 2.9 & 5.4 & $<$ 62\\
HD 163296 & 0.55 $\times$ 0.47 & 76.8 & 2.5 & 4.9 & 4396 $\pm$ 440\\
IM Lup & 0.56 $\times$ 0.55 & 2.3 & 8.0 & 12.9 & 1450 $\pm$ 170\\
LkCa 15 & 0.59 $\times$ 0.49 & -26.3 & 3.0 & 5.5 & 2435 $\pm$ 250\\
MWC 480 & 0.73 $\times$ 0.47 & 13.6 & 4.2 & 7.7 & 1846 $\pm$ 190\\
HD 143006 & 0.58 $\times$ 0.46 & 87.9 & 5.8 & 9.7 & 435 $\pm$ 56\\
J1604-2130 & 0.56 $\times$ 0.45 & -87.9 & 5.8 & 8.3 & 2632 $\pm$ 270\\
J1609-1908 & 0.56 $\times$ 0.45 & -86.5 & 5.8 & 10.8 & 563 $\pm$ 70\\
J1612-1859 & 0.55 $\times$ 0.45 & -87.4 & 5.6 & 11.1 & $<$ 97\\
J1614-1906 & 0.56 $\times$ 0.45 & -86.8 & 5.6 & 11.3 & $<$ 100\\
V4046 Sgr & 0.97 $\times$ 0.70 & 83.2 & 4.0 & 8.1 & 2962 $\pm$ 300\\
\enddata
\tablenotetext{}{$^a$For 0.5 km s$^{-1}$ channel widths.  $^b$Median rms; see Section \ref{sec:datared}.  $^c$3$\sigma$ upper limits are reported for nondetections.  Uncertainties are derived by bootstrapping, added in quadrature with a 10\% calibration uncertainty.  $^d$ SMA observation.  $^e$Tentatitve detection.}
\end{deluxetable*}

The moment zero rms is also estimated by bootstrapping.  In the Keplerian mask-extracted moment zero maps, each pixel is the sum of a different number of channels, and therefore the rms varies across the map.  The rms for each moment zero map pixel is calculated from the standard deviation of flux values in the same pixel location in 1000 emission-free masked images.  The median pixel rms across the map is used to represent the moment zero rms in Table \ref{tab:fluxes_hcn_c2h} and Figure \ref{fig:mom0}.  

\section{Observational results}
\label{sec:obs_res}
\begin{figure*}
	\includegraphics[width=\linewidth]{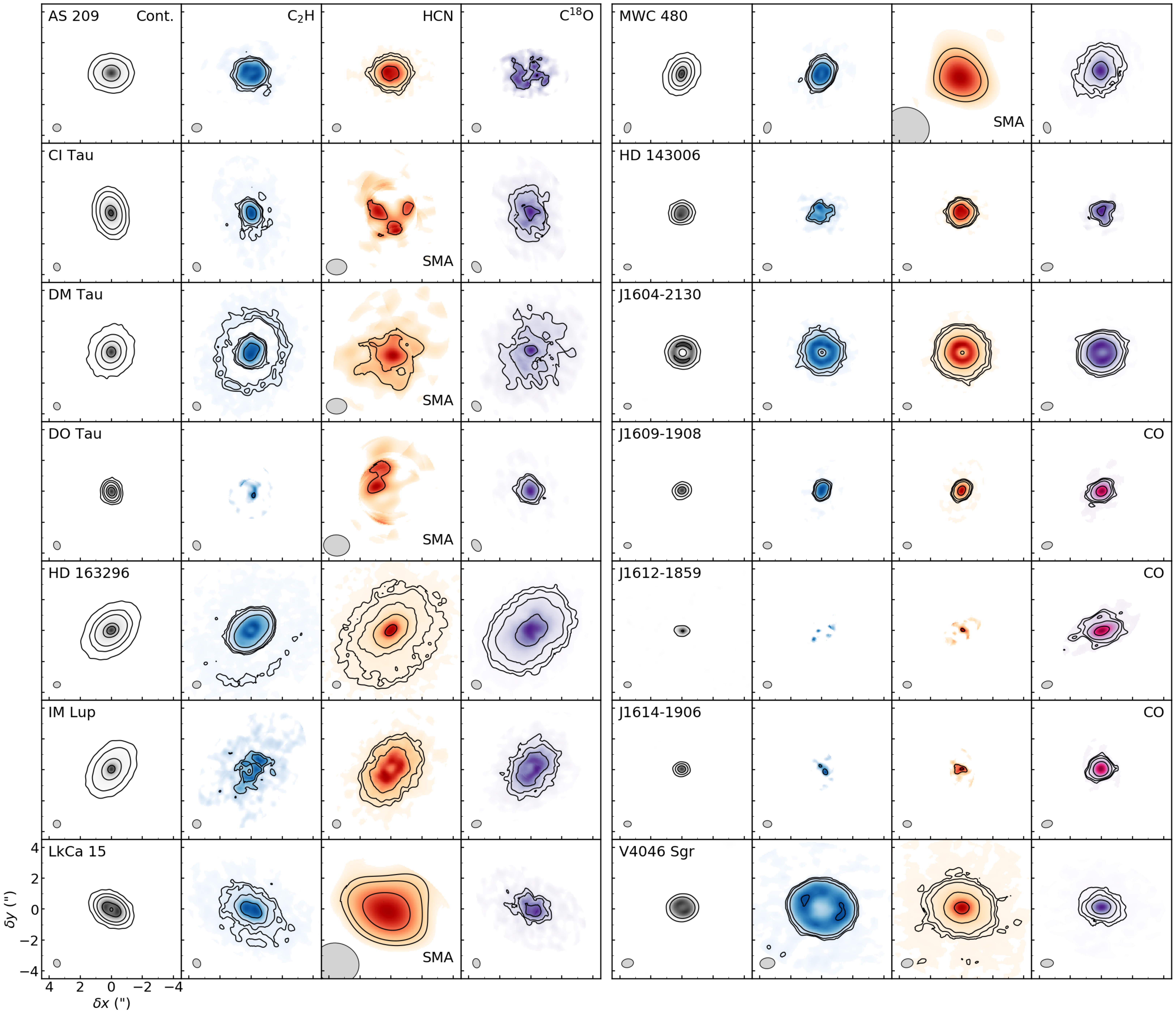}
	\caption{First column shows the 1.1mm dust continuum maps for each disk, with contours drawn at 5, 30, 100, 400, and 800$\sigma$.  Second, third, and fourth columns show moment zero maps for C$_2$H N = 3 --2, J = $\frac{7}{2}$ -- $\frac{5}{2}$; HCN J = 3 -- 2; and C$^{18}$O or CO J = 2 -- 1, respectively.  Line maps were extracted with Keplerian masking, and contour levels show 3, 5, 10, 30, and 100$\times$ the median rms.  Top panels indicate the line shown in each column, except CO where indicated.  Synthesized beams are shown in the lower left of each panel.  Color scales are normalized to each individual line image.  We note that HCN observations of CI Tau, DM Tau, DO Tau, LkCa 15, and MWC 480 were taken with the SMA, and all other observations with ALMA.}
\label{fig:mom0}
\end{figure*}

Figure \ref{fig:mom0} shows the 1.1mm dust continuum as well as line emission moment zero maps for the C$_2$H J = $\frac{7}{2}$ -- $\frac{5}{2}$, HCN J = 3 -- 2 and C$^{18}$O or CO J = 2 -- 1 transitions.  Figure \ref{fig:radprof} shows the deprojected and azimuthally averaged radial profiles for all line observations.  Moment zero maps for the C$_2$H J = $\frac{5}{2}$ -- $\frac{3}{2}$ are shown in Appendix \ref{sec:app_obsdat}, along with disk-integrated line spectra.  Based on these results, we present molecule detections and emission morphologies for this sample.

\subsection{Molecule detections}
We consider a line detected if the integrated flux density is significant above a 3$\sigma$ level and at least 3 channels show emission $>3\times$ rms within the Keplerian mask.  Based on these criteria, we detect C$_2$H and HCN towards all disks except DO Tau, J1612-1859, and J1614-1906.  HCN is also tentatively detected towards CI Tau: although the integrated flux is $>3\sigma$, it does not appear in the individual channel maps.  Because it is unclear if the HCN emission in CI Tau is real, we treat it as a non-detection for subsequent analysis.  C$^{18}$O is detected towards all disks except J1609-1908, J1612-1859, and J1614-1906, for which we instead use the major isotopologue $^{12}$C$^{16}$O (CO) for subsequent analysis.

\subsection{Emission morphologies}
HCN emission is centrally peaked towards all disks except AS 209, IM Lup, and J1604-2130.  In all disks with detections, HCN is more extended than the dust continuum.  We note that HCN observations taken with the SMA are not well resolved and there may be sub-structure in the HCN emission that is hidden in the moment zero maps presented here.  

C$^{18}$O is also centrally peaked towards all disks other than AS 209, IM Lup, LkCa 15, and J1604-2130.  In most cases, the C$^{18}$O profile shows a weak shoulder extending out past the continuum edge in addition to strong emission within the dust continuum.  

C$_2$H emission exhibits a variety of morphologies: it is centrally depressed in AS 209, HD 163296, IM Lup, MWC 480, HD 143006, J1604-2130, and V4046 Sgr, and centrally peaked in the remaining disks with detections.  Additionally, in DM Tau, HD 163296, and V4046 Sgr, the C$_2$H moment zero maps  shows a clear ring past the dust continuum.  The deprojected radial profiles of AS 209, CI Tau, IM Lup, and LkCa 15 show a weak shoulder extending out beyond the dust continuum, but no sharp ringed feature.  MWC 480, HD 143006, J1604-2130, and J1609-1908 show more compact morphologies, though for disks with fainter emission we may be sensitivity-limited in detecting a weak shoulder feature.

J1604-2130 shows a central gap for most molecular emission \citep[see also][]{vanderMarel2015}, which is likely related to the presence of a large dust cavity \citep[e.g.][]{Zhang2014, Dong2017}.  The central emission dips for all molecules in IM Lup could be a dust opacity effect \citep{Cleeves2016}.  IM Lup is also a young, cold, and massive disk, which may affect its chemistry relative to the other disks in the sample.  

\begin{figure*}
\begin{centering}
	\includegraphics[width=\linewidth]{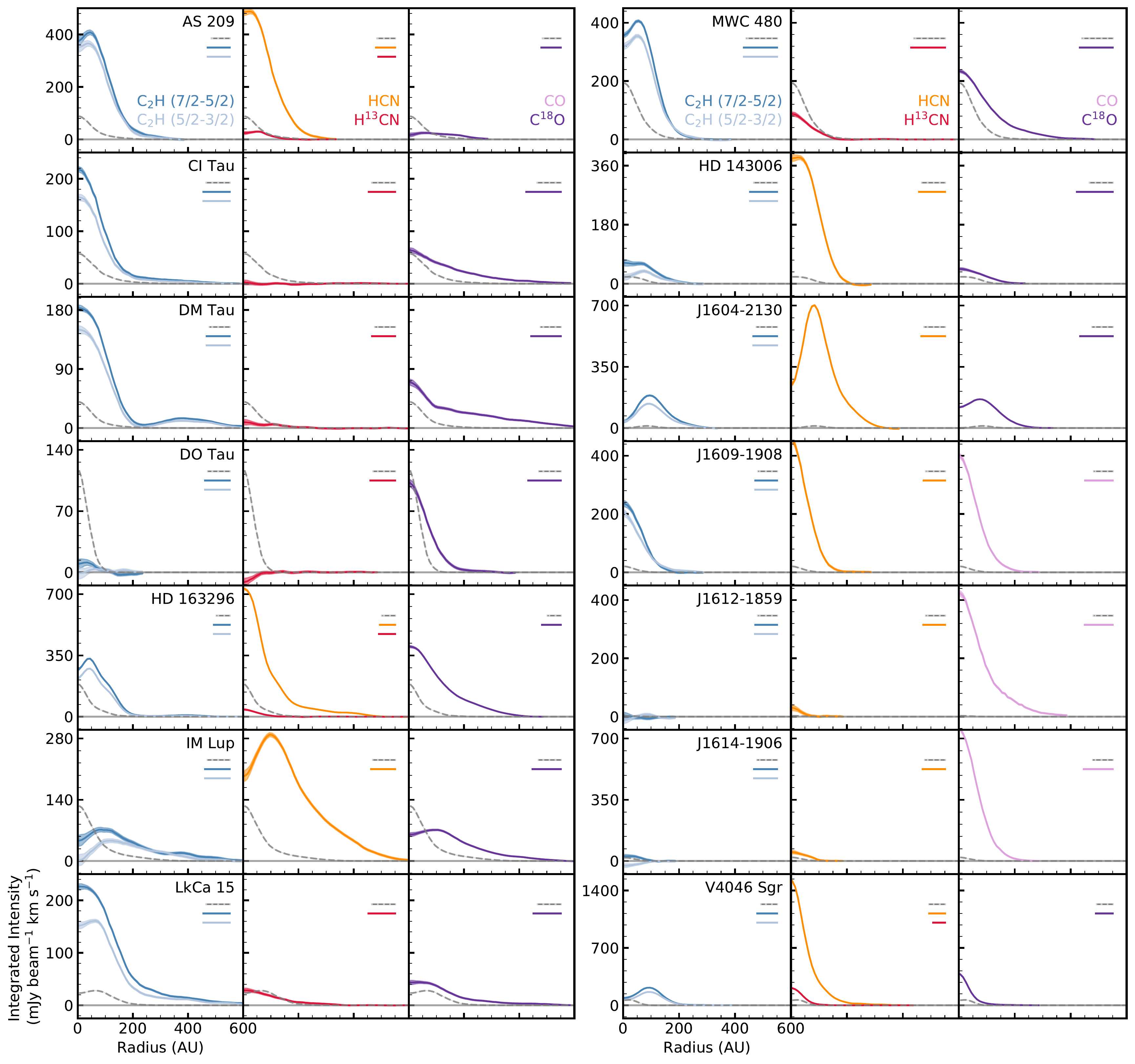}
	\caption{Deprojected and azimuthally averaged intensity profiles for C$_2$H N = 3 -- 2, J = $\frac{7}{2}$ -- $\frac{5}{2}$ and J = $\frac{5}{2}$ -- $\frac{3}{2}$ transitions (left); HCN and H$^{13}$CN J = 3 -- 2 transitions (middle); and C$^{18}$O and CO J = 2 -- 1 transitions (right).  The continuum profile (dashed grey) is shown in each panel for reference, with y axis units of mJy beam$^{-1}$.  Bars in the upper right of each panel represent the restoring beam major axes.  HCN observations taken with the SMA are not shown.}
	\end{centering}
\label{fig:radprof}
\end{figure*}

\section{Optical depth \& column density derivation}
\label{sec:tau_cd}
In the following sections we describe the use of hyperfine fitting and isotopologue ratios to derive C$_2$H and HCN optical depths.  For disks with optical depth constraints on C$_2$H or HCN, we also measure excitation temperatures and column densities; for disks without direct optical depth measurements, we estimate column densities assuming reasonable excitation temperatures and optical depths.

\subsection{Hyperfine fitting}
\label{sec:hf}
Both the C$_2$H and HCN transitions covered in this survey exhibit hyperfine splitting.  By fitting hyperfine-split spectra we can derive line optical depths and excitation temperatures, and in turn molecular column densities.  In general, the disk-integrated spectra of our sample suffer too much line blending to enable hyperfine fitting, however for sources with bright emission we are able to fit the spectra in individual pixels, which have much narrower line widths.  Because of the close spacing of the HCN hyperfine lines, the only disk for which we could fit the hyperfine structure is J1604-2130, a nearly face-on disk with very narrow lines.  The C$_2$H hyperfine components are more widely spaced, and the following disks had sufficient signal to noise to fit: AS 209, CI Tau, DM Tau, HD 163296, J1604-2130, LkCa 15, MWC 480, and V4046 Sgr.  Spectral line parameters used to fit the hyperfine-resolved transitions are listed in Table \ref{tab:linedat}.

\begin{deluxetable*}{lrrrrrrrrr} 
	\tabletypesize{\footnotesize}
	\tablecaption{Spectral line data \label{tab:linedat}}
	\tablecolumns{10} 
	\tablewidth{\textwidth} 
	\tablehead{
		\colhead{Molecule}           &
		\colhead{Transition}         &
		\colhead{Frequency}       &
		\colhead{S$\mu^2$}        &
		\colhead{$g_u$}              & 
		\colhead{$E_u$}             &
		\multicolumn{2}{c}{$Q(T)$}  & 
		\colhead{$R_i$}                &
		\colhead{$R_m$}             \\
		\colhead{}                        & 
		\colhead{(GHz)}               & 
		\colhead{}                        & 
		\colhead{(D$^2$)}            &
		\colhead{}                        &
		\colhead{(K)}                   &
		\colhead{300 K}                        & 
		\colhead{9.375 K}                        & 
		\colhead{}                        & 
		\colhead{}                        
		}
\startdata
\multicolumn{10}{c}{Hyperfine-resolved transitions$^a$} \\
\hline
C$_2$H$^b$ & N = 3 -- 2, J = $\frac{7}{2} - \frac{5}{2}$, F = 4 -- 3 &262.00422660  &  2.47  &  9  &  25.15  & 574.2 & 19.3 &  1.0 & 0.572 \\
& N = 3 -- 2, J = $\frac{7}{2} - \frac{5}{2}$, F = 3 -- 2 & 262.00640340  & & & &&& 0.748 & \\
& N = 3 -- 2, J = $\frac{5}{2} - \frac{3}{2}$, F = 3 -- 2 & 262.06484330 & & &  &&& 0.714& \\
& N = 3 -- 2, J = $\frac{5}{2} - \frac{3}{2}$, F = 2 -- 1 & 262.06733120  & & & &&& 0.467& \\
\hline
HCN$^c$ & J = 3 --2, F = 3 -- 3 & 265.88489120 & & &&& &0.086 &\\
 & J = 3 --2, F = 2 -- 1 & 265.88618860 & &&&&& 0.467 &\\
 & J = 3 --2, F = 3 -- 2 & 265.88643390 &&&&&&0.691 &\\
 & J = 3 --2, F = 4 -- 3 & 265.88649990 & 34.37 & 9 & 25.52 &453.5 & 14.3 & 1.0 & 0.429 \\
 & J = 3 --2, F = 2 -- 3 & 265.88697930 &&&&&&0.002  &\\
 & J = 3 --2, F = 2 -- 2 & 265.88852210 &&&&&& 0.086 &\\
 \hline
 \multicolumn{10}{c}{Hyperfine-unresolved transitions} \\
 \hline
C$_2$H$^b$ & N = 3 -- 2, J = $\frac{7}{2} - \frac{5}{2}$ & 262.00422660 & 4.3 & 16 & 25.15 & 574.2 & 19.3  & & \\
HCN$^c$ & J = 3 -- 2 & 265.88643390 & 80.2 & 21 & 25.52 &453.5 & 14.3& & 
\enddata
\tablenotetext{}{$^a$Parameters required for column density calculations are shown for the main hyperfine line. $^b$ From the JPL catalogue. $^c$ From the CDMS catalogue.}
\end{deluxetable*}

Our fitting procedure is adapted from \citet{Estalella2017} and \citet{Mangum2015}.  The optical depth $\tau_k$ in a velocity channel $k$ is given by:
\begin{equation}
\tau_k = \sum_{i=1}^{N_{hf}}\frac{\tau_m R_i}{\Delta V_{ch}}G.
\end{equation}
Here $N_{hf}$ is the total number of hyperfine transitions, $R_i$ is the intensity of each hyperfine line relative to the main line, $\Delta V_{ch}$ is the channel width, and $G$ describes the line profile:
\begin{equation}
G = \frac{\sqrt{\pi}}{4\sqrt{\ln2}}{\Delta V_i}[\rm{erf}(x^+) - \rm{erf}(x^-)],
\end{equation}
with
\begin{equation}
x^\pm = 2\sqrt{\ln2} \frac{V_k \pm \Delta V_{ch}/2 - V_{lsr} - V_{i}}{\Delta V_i}.
\end{equation}
$V_k$ is the central velocity of channel $k$, $V_{lsr}$ is the line center of the main hyperfine line, and $\Delta V_i$ is the velocity offset of each hyperfine line to the main line.

Given the Rayleigh-Jeans equivalent temperature:
\begin{equation}
J_\nu(T) = \frac{hv/k_b}{\mathrm{exp} (hv/k_bT)-1},
\end{equation}
the model hyperfine spectrum is written as:
\begin{equation}
T_R = f[J_\nu(T_{ex}) - J_\nu(T_{bg})](1-e^{-\tau_\nu}),
\label{eq_Tr}
\end{equation}
where $T_R$ Is the source brightness temperature; $f$ is the filling factor, equal to 1 for spatially resolved emission; $T_{ex}$ and $T_{bg}$ are the excitation temperature and background temperature respectively; and $\tau_\nu$ is the optical depth.  We define the following parameters $A$, $A_m^*$, and $\tau_m^*$ as:
\begin{eqnarray}
A = f[J_\nu(T_{ex}) - J_\nu(T_{bg})] \\
A_m^* = A(1-e^{-\tau_m}) \\
\tau_m^* = 1-e^{-\tau_m}
\end{eqnarray}
We simultaneously fit each spectrum for the parameters $\Delta V_i$, $V_{lsr}$, $A_m^*$, and $\tau_m^*$.  We note that this fitting assumes LTE conditions; see Section \ref{sec:emission_origin} for a more detailed discussion of the excitation conditions in our sample.  The total optical depth of the line can then be determined from:
\begin{equation}
\tau = \tau_m / R_m,
\end{equation}
where $R_m$ is the fractional strength of the main hyperfine line relative to other F-level transitions for a given J-level transition.  The excitation temperature can then be solved for using equation \ref{eq_Tr}.  Finally, we solve for the total column density $N_T$:
\begin{multline}
N_T = \frac{3h}{8\pi^3S\mu^2} \frac{Q(T_{ex})}{g_u} 
\frac{\mathrm{exp}(E_u/T_{ex})}{\mathrm{exp}(h\nu/k_bT_{ex})-1}  \\ 
\times \frac{\int T_R dV}{f[J_\nu(T_{ex}) - J\nu(T_{bg})]} \frac{\tau}{1-e^{-\tau}},
\label{eq:Nt}
\end{multline}
where $S\mu^2$ is the line intensity, $Q(T_{ex})$ is the partition function, $g_u$ is the upper state degeneracy, and $E_u$ is the upper state energy.

In practice, we extract spectra from each pixel within a slice along the disk major axis the width of the beam major axis.  In this way the derived radial profiles are less impacted by disk inclination effects on the optical depth compared to pixels with large offsets from the disk major axis.  All pixels with a peak SNR $<$ 6 are excluded from fitting to ensure sufficient signal to noise for the fitting procedure.  Spectra are converted from flux density $S_\nu$ to brightness temperature by:
\begin{equation}
T_R = \frac{S_\nu c^2}{2k_b\nu^2\Omega},
\end{equation}
where $\Omega$ is the angular size of each pixel.  The background temperature $T_{bg}$ is taken from the continuum brightness temperature in the same pixel.  We use the affine-invariant MCMC package {\fontfamily{qcr}\selectfont emcee} \citep{Foreman-Mackey2013} to sample posterior distributions of the four fit parameters.  Example hyperfine fits for each source can be found in Appendix \ref{sec:hf_ex}.

We exclude from further analysis any pixels with derived line widths greater than 2.5 km s$^{-1}$ since line blending is too severe to obtain good constraints on the fit parameters.  Likewise, we exclude pixels with derived optical depths outside the range of 0.15 to 30, since the spectrum is not sensitive to changes in the optical depth beyond this.

\begin{figure*}
\begin{centering}
	\includegraphics[width=0.85\linewidth]{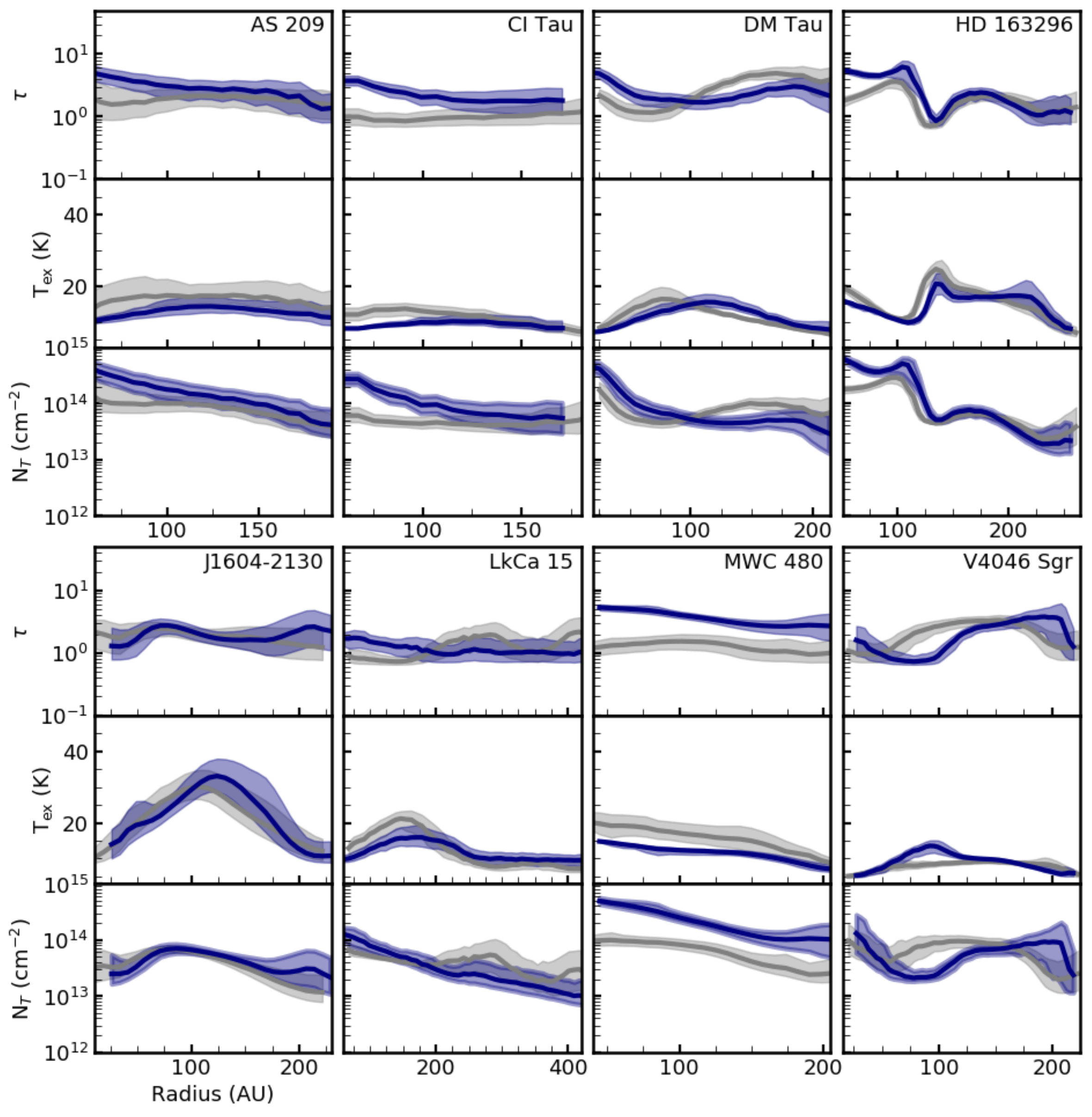}
	\caption{Deprojected and radially averaged C$_2$H optical depth, excitation temperature, and column density profiles derived from hyperfine fitting.  Lines represent the median of the down-sampled and pooled posteriors of all pixels in each radial bin, and shaded regions show 2$\sigma$ confidence intervals.  Purple and grey represent projections north and south of the source center along the disk major axis.}
\end{centering}
\label{fig:taurad_c2h}
\end{figure*}

Figure \ref{fig:taurad_c2h} shows the optical depth, excitation temperature, and column density profiles derived from C$_2$H hyperfine fitting.  The hyperfine fits projecting in opposite directions from the source center (blue and grey in Figure \ref{fig:taurad_c2h}) are in most cases quite similar, but in some cases (e.g. CI Tau, MWC 480) show variations indicative of azimuthal asymmetry in the disk.  In most cases, C$_2$H emission is marginally optically thick, with a $\tau$ just around or above unity.  C$_2$H excitation temperatures are typically between $\sim$10 and 40K, with column densities around 10$^{13}$ -- 10$^{14}$ cm$^{-2}$.

\subsection{Isotopologue ratios}
\label{sec:isotope}
\begin{figure}
	\includegraphics[width=\linewidth]{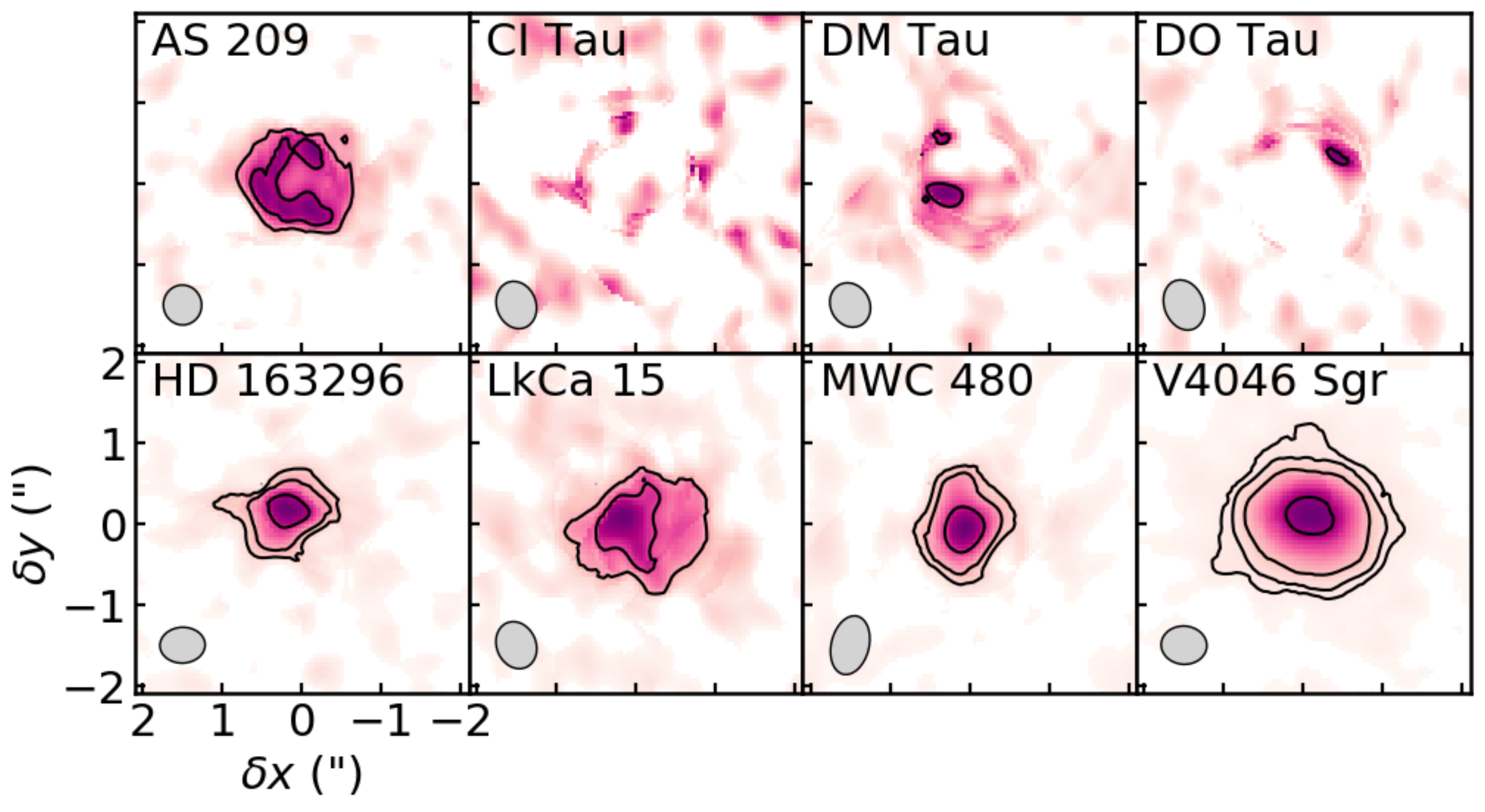}
	\caption{Moment zero maps for the H$^{13}$CN J = 3 -- 2 transition, extracted with Keplerain masking.  Contour levels show 3, 5, 10, 30, and 100 $\times$ the median rms.  Restoring beams are shown in the lower left of each panel.}
\label{fig:mom0_h13cn}
\end{figure}

For disks with H$^{13}$CN detections (AS 209, HD 163296, V4046 Sgr, LkCa 15, and MWC 480), we can obtain estimates of the HCN optical depth from the HCN/H$^{13}$CN intensity ratio.  Moment zero maps for all H$^{13}$CN observations are shown in Figure \ref{fig:mom0_h13cn}.  As described in \citet{Mangum2015}, assuming co-spatial emission of both molecules and that the molecular $^{12}$C/$^{13}$C ratio is the same as the atomic ratio:
\begin{equation}
\frac{T_R(\mathrm{HCN})}{T_R(\mathrm{H^{13}CN})} \approx 
\frac{1-\mathrm{exp}(-\tau_{\nu,\mathrm{HCN}})}{1-\mathrm{exp}(-f(\frac{\mathrm{^{13}C}}{\mathrm{^{12}C}})\tau_{\nu,\mathrm{HCN}})},
\end{equation}
where $f(\mathrm{^{13}C}/\mathrm{^{12}C)}$ is the atomic  $^{13}$C/$^{12}$C ratio.  We adopt a $^{12}$C/$^{13}$C value of 68 $\pm$ 15 appropriate for the local ISM \citep{Milam2005}.

Ideally this is done using the ratio of brightness temperatures at the line center for each transition.  However, due to low H$^{13}$CN signal-to-noise, we instead use the ratio of velocity-integrated intensities, making the assumption that HCN and H$^{13}$CN have similar line profiles:
\begin{equation}
\frac{T_R(\mathrm{HCN})}{T_R(\mathrm{H^{13}CN})} \approx  \frac{\int T_R(\mathrm{HCN})dV}{\int T_R(\mathrm{H^{13}CN})dV}.
 \end{equation}
Also due to low signal to noise, we cannot perform this analysis for individual pixels, but instead estimate optical depths in radial bins.  We construct radially averaged deprojected intensity profiles for HCN and H$^{13}$CN including only pixels in which the H$^{13}$CN signal to noise ratio is greater than 2.5, to ensure that the isotopologue ratio is not diluted by non-detections.  From the radial $\tau_{\mathrm{HCN}}$ profiles we then estimate $T_{ex}$ and $N_T$ as for C$_2$H, using equations \ref{eq_Tr} and \ref{eq:Nt} along with the hyperfine-unresolved spectral line parameters listed in Table \ref{tab:linedat}.

We note that LkCa 15 and MWC 480 HCN observations were taken with the SMA and have much larger beams than the ALMA observations.  We taper the H$^{13}$CN observations in these sources to roughly match the SMA beam sizes, enabling a more direct comparison of flux ratios.  Beam dilution is likely an issue for these sources since the emission is not well resolved.

\begin{figure*}
\begin{centering}
	\includegraphics[width=\linewidth]{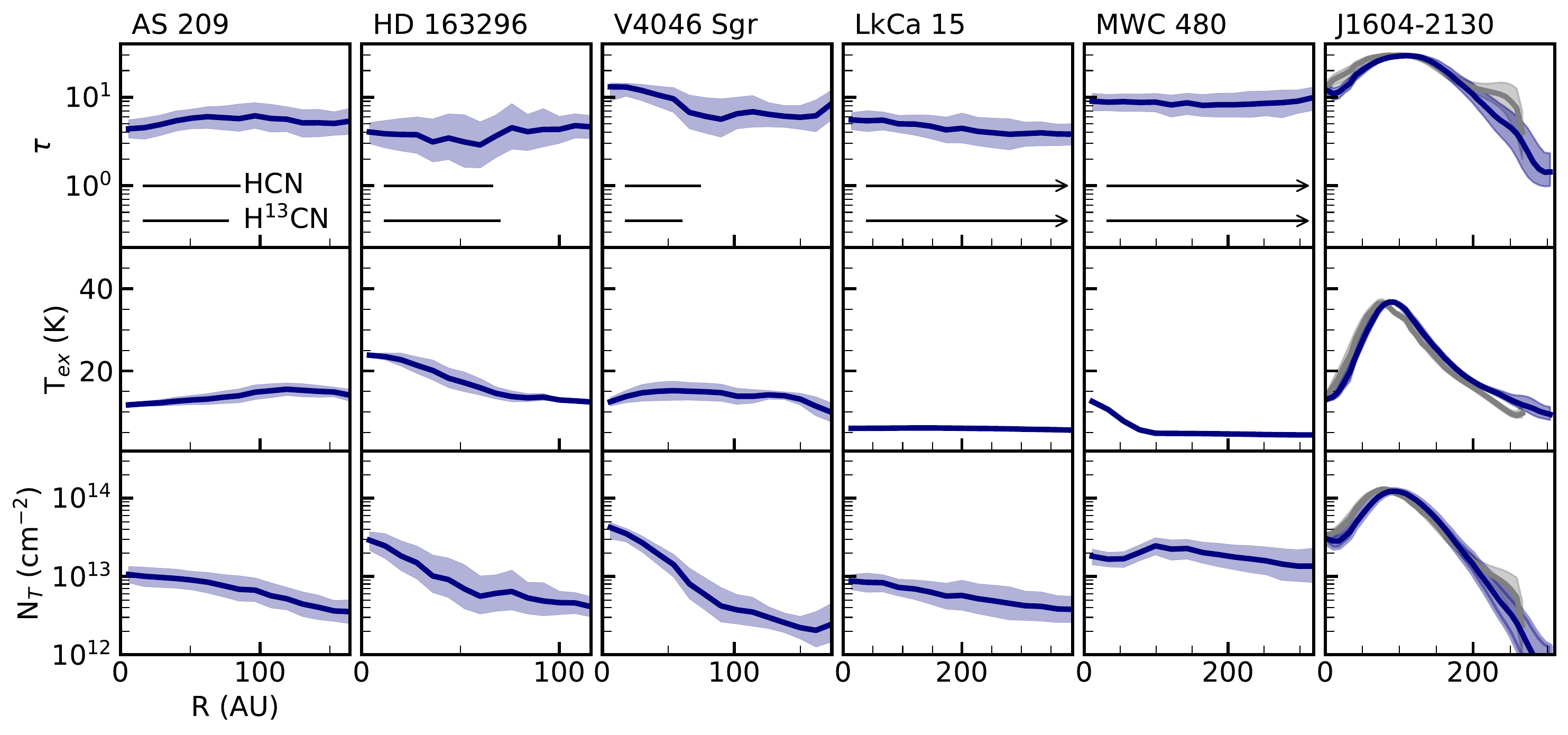}
	\caption{Deprojected and radially averaged HCN optical depth, excitation temperature, and column density profiles.  For AS 209, HD 163296, V4046 Sgr, LkCa 15, and MWC 480, optical depths are derived from the H$^{13}$CN isotopologue ratio.  Shaded regions show uncertainties based on the standard deviations in the HCN/H$^{13}$CN ratio and the HCN brightness temperature in each radial bin, as well as the uncertainty in the $^{12}$C/$^{13}$C ratio.  The beam major axis for HCN and H$^{13}$CN are shown in the top panel for each source; arrows are used when the beam is larger than the x axis range.  For J1604-2130, profiles are derived from HCN hyperfine fitting.  Lines represent the median of the down-sampled and pooled posteriors of all pixels in each radial bin, and shaded regions show 2$\sigma$ confidence intervals.  Blue and gray represent projections north and south of the source center along the disk major axis.}
	\end{centering}
\label{fig:taurad_hcn}
\end{figure*}

Figure \ref{fig:taurad_hcn} shows the optical depth, excitation temperature, and column density profiles derived for HCN using isotopologue ratios and, in J1604-2130, hyperfine fitting.  In our sample HCN is somewhat optically thick, with $\tau$ typically between 3 and 10.  Excitation temperatures are typically between $\sim$10 and 20K, and the very low temperatures found for MWC 480 and LkCa 15 are likely due to beam dilution.  For J1604-2130, the excitation temperatures are somewhat higher than the rest of the disk sample.  We note that overlap in the HCN hyperfine components is known to lead to anomalous intensity ratios \citep[e.g.][]{Gonzalez1993, Daniel2008}, and in particular radiative pumping of the J = 3 level can lead to increased excitation temperatures for the J = 3 -- 2 transition \citep{Magalhaes2018}.  Detailed radiative transfer modeling is required to evaluate if the warmer temperature in J1604-2130 is real or results from the hyperfine fitting technique.  Still, while our HCN hyperfine fits slightly under-estimate the intensity of the blended F = 2 --1, F = 3 -- 2, and F = 4 -- 3 feature, the overall fit is quite good (Appendix \ref{sec:hf_ex}), and we expect that any hyperfine anomaly is very minor in this case.  In general, the optical depths and column densities obtained from HCN hyperfine fitting for J1604-2130 are comparable to the isotopologue results for other disks, and this agreement points to the robustness of both methods.

\subsection{Disk-averaged optical depths, excitation temperatures, and column densities}
\label{sec:est_cd}
Disk-averaged column densities are calculated from Equation \ref{eq:Nt} using the temperature and optical depth constraints from Sections \ref{sec:hf} and \ref{sec:isotope}, the integrated line fluxes from Table \ref{tab:fluxes_hcn_c2h}, and the unresolved spectral line parameters in Table \ref{tab:linedat}.  To avoid severe beam dilution, for each molecule we calculate the angular size $\Omega$ assuming a radial extent within which 50\% of the line emission is contained.

There are two major uncertainties in these calculations: excitation temperatures and line opacities.  To address the former we calculate column densities for 10K and 30K excitation temperatures, encompassing the range found in Sections \ref{sec:hf} and \ref{sec:isotope}.  We address the opacity uncertainties by similarly calculating column densities for two extremes: with no optical depth correction, i.e. a column density lower limit; and corrected for an assumed optical depth of 2 for C$_2$H and 8 for HCN, typical opacities for these molecules at small radii based on our hyperfine and isotopologue analysis.  

The resulting column densities are listed in Table \ref{tab:Nts}.  For a single source, we obtain the lowest column density estimate assuming 30K emission without optical depth correction and the highest estimate assuming 10K emission and an optical depth $>$1.  For C$_2$H, this range of estimates varies by a factor of 4 -- 5.  HCN estimates are more sensitive to our assumptions, varying by a factor of 15 -- 17.  Assuming 10K excitation temperatures and optically thick emission, column density estimates across the source sample range from 5 -- 37 $\times$ 10$^{12}$ cm$^{-2}$ for C$_2$H, and from 1 -- 10 $\times$ 10$^{12}$ cm$^{-2}$  for HCN.

\begin{deluxetable*}{lccccccc} 
	\tabletypesize{\footnotesize}
	\tablecaption{C$_2$H and HCN radial extents and disk-averaged column density estimates \label{tab:Nts}}
	\tablecolumns{8} 
	\tablewidth{\textwidth} 
	\tablehead{
		\colhead{Source}           &
		\colhead{R$_{\mathrm{C_2H}}$ (AU)} &
		\multicolumn{2}{c}{N(C$_2$H) (10$^{12}$ cm$^{-2}$)}         &
		\colhead{} &
		\colhead{R$_{\mathrm{HCN}}$ (AU)} &
		\multicolumn{2}{c}{N(HCN) (10$^{12}$ cm$^{-2}$)}             \\      
		\colhead{}  &
		\colhead{} &
		\colhead{10K} & 
		\colhead{30K} &
		\colhead{ } &
		\colhead{} &
		\colhead{10K} & 
		\colhead{30K} 
		}
\startdata
AS 209 & 275 & 18.2 [7.9] & 9.7 [4.2]  &  & 235 & 3.5 [0.4] & 1.8 [0.2] \\
CI Tau & 415 & 5.3 [2.3] & 2.8 [1.2]  &  & -- & -- & -- \\
DM Tau & 485 & 6.3 [2.7] & 3.3 [1.4]  &  & 485 & 1.0 [0.1] & 0.5 [0.1] \\
HD 163296 & 485 & 6.7 [2.9] & 3.6 [1.5]  &  & 485 & 1.4 [0.2] & 0.8 [0.1] \\
IM Lup & 485 & 5.4 [2.3] & 2.9 [1.2]  &  & 515 & 1.9 [0.2] & 1.0 [0.1] \\
LkCa 15 & 515 & 8.0 [3.5] & 4.3 [1.9]  &  & 485 & 1.5 [0.2] & 0.8 [0.1] \\
MWC 480 & 275 & 22.2 [9.6] & 11.9 [5.1]  &  & 415 & 1.0 [0.1] & 0.5 [0.1] \\
HD 143006 & 205 & 9.9 [4.3] & 5.3 [2.3]  &  & 205 & 4.9 [0.6] & 2.5 [0.3] \\
J1604-2130 & 235 & 37.3 [16.1] & 19.9 [8.6]  &  & 275 & 10.4 [1.3] & 5.4 [0.7] \\
J1609-1908 & 205 & 8.8 [3.8] & 4.7 [2.0]  &  & 205 & 1.7 [0.2] & 0.9 [0.1] \\
V4046 Sgr & 275 & 7.1 [3.1] & 3.8 [1.6]  &  & 305 & 2.0 [0.3] & 1.1 [0.1] \\
\enddata
\tablenotetext{}{First listed numbers are corrected for an assumed optical depth of 2 for C$_2$H and 8 for HCN; bracketed numbers are uncorrected (i.e., column density lower limits).}
\end{deluxetable*}

\section{Disk modeling}
\label{sec:diskmodels}
In our disk sample we observe a wide range in C$_2$H emission morphologies, including centrally peaked vs. centrally depressed emission, compact vs. extended emission, and the presence vs. absence of emission rings external to the dust continuum.  This degree of morphological diversity is not seen for the CO or HCN molecular lines observed in our sample.  In this section we explore how varying physical properties across the disk sample can explain the range of observed C$_2$H morphologies.

Several pathways could be responsible for C$_2$H formation in disks.  Based on disk chemical modeling in \citet{Bergin2016}, the main channels to C$_2$H are predicted to be:
\begin{equation}
\label{eq_C0}
\begin{aligned}
\mathrm{C + H_3^+ \rightarrow CH_3^+} \\
\mathrm{CH_3^+ + H_2 \rightarrow CH_5^+ }\\
\mathrm{CH_5^+} + e^- \mathrm{\rightarrow CH_3 + H_2 }\\
\mathrm{CH_3 + C \rightarrow C_2H_2 }\\
\mathrm{C_2H_2} + h\nu \rightarrow \mathrm{C_2H + H }\\
\mathrm{C_2H_2 + H_3^+ \rightarrow C_2H_3^+ + H_2 }\\
\mathrm{C_2H_3^+} + e^- \mathrm{\rightarrow C_2H + H_2}
\end{aligned}
\end{equation}
\noindent C$_2$H formation is initiated by neutral atomic carbon and H$_3^+$, which in disk atmospheres forms from X-ray ionization of H$_2$, followed by H$_2^+$ reaction with H$_2$.  The final step to form C$_2$H in this scheme depends on either photodissociation of C$_2$H$_2$ or recombination of C$_2$H$_3^+$ with an electron.  In both cases, UV radiation is required since the dominant source of electrons in disk atmospheres is C photoionization.

Another possible channel to C$_2$H formation drawn from PDR chemical models \citep{Federman1994, Nagy2015} instead involves ionized atomic carbon reacting with H$_2$:
\begin{equation}
 \label{eq_Cpos}
\begin{aligned}
\mathrm{C} +  h\nu \rightarrow \mathrm{C^+} + e^-\\
\mathrm{C^+ + H_2 \rightarrow CH_2^+} \\
\mathrm{CH_2^+} + e^- \mathrm{\rightarrow CH + H} \\
\mathrm{CH + C^+ \rightarrow C_2^+ + H} \\
\mathrm{C_2^+ + H_2 \rightarrow C_2H^+ + H} \\
\mathrm{C_2H^+ + H_2 \rightarrow C_2H_2^+ + H} \\
\mathrm{C_2H_2^+} + e^- \mathrm{\rightarrow C_2H + H}
\end{aligned}
\end{equation}
\noindent Here, UV radiation is key both for initiating the chemistry through C$^+$ formation, and for generating the electrons responsible for intermediate and final product formation.  This pathway may be especially relevant in the heavily UV-irradiated disk atmosphere, and indeed recent carbon isotopologue measurements point to a C$^+$ driven C$_2$H formation pathway in disks (E. Bergin, in preparation).

Schemes \ref{eq_C0} and \ref{eq_Cpos} both depend on the availability of atomic carbon and UV radiation to form C$_2$H.  We aim to test whether these factors can explain the range of C$_2$H morphologies observed in our disk sample.  While scheme \ref{eq_C0} additionally requires X-ray photons to initiate H$_3^+$ formation, for an initial exploration of the physical properties driving C$_2$H formation we focus on the factors common to both pathways.  In the following sections we explore how variations in disk physical properties impact the local UV fluxes and carbon atom abundances by constructing physical models first for generic toy disk models, and then for source-specific models of a subset of the disks in our sample.

\subsection{Disk model setup}
\label{sec:app_model}
Our physical model is based on the formalism presented in \citet{Andrews2011} and \citet{Rosenfeld2013} for a simple accretion disk \citep[see also][]{Lynden1974, Hartmann1998}.  Dust surface density profiles $\Sigma_{dust}$ are parameterized as:
\begin{equation}
\Sigma_{dust}(r) = \Sigma_c \Big{(}\frac{r}{R_c}\Big{)}^{-\gamma} \mathrm{exp} \Big{[}\Big{(}\frac{r}{r_c}\Big{)}^{2-\gamma} \Big{]},
\end{equation}
where $\Sigma_c$ is the surface density at the characteristic radius $R_c$ and $\gamma$ is the viscosity power-law index.  The volumetric dust density is calculated for two separate dust populations: large grains or pebbles in the settled midplane, and small grains extending into the disk atmosphere.  Both midplane and atmosphere grain populations begin at an inner radius $R_{\mathrm{in}}$ and extend to a radius $R_{\mathrm{peb}}$ and $R_{\mathrm{out}}$ respectively, where $R_{\mathrm{peb}} < R_{\mathrm{out}}$ to mimic settling and drift of pebbles.  We set  $R_{\mathrm{in}}$ to 0.1AU for all models.  The volumetric density of each grain population is assumed to fall off as a Gaussian with increasing disk height $z$:
\begin{equation}
\rho_{dust}(r,z) = X_i\frac{\Sigma_{dust}(r)}{\sqrt{2\pi}H_i(r)}\mathrm{exp}\Big{[}-0.5\Big{(}\frac{z}{H_i(r)}\Big{)}^2\Big{]},
\end{equation}
where $X_i$ is the density fraction of each grain population $i$ (assumed to be 0.9 for midplane grains and 0.1 for atmospheric grains), and $H_i$ is the scale height of each population:
\begin{eqnarray}
H_{atm}(r) = H_c \Big{(}\frac{r}{R_H}\Big{)}^h \\
H_{mid}(r) = 0.25 H_{atm}(r)
\end{eqnarray}
where $H_c$ is the scale height at the characteristic radius $R_H$, and $h$ is the scale height power law index.

\begin{figure}
\begin{centering}
	\includegraphics[width=\linewidth]{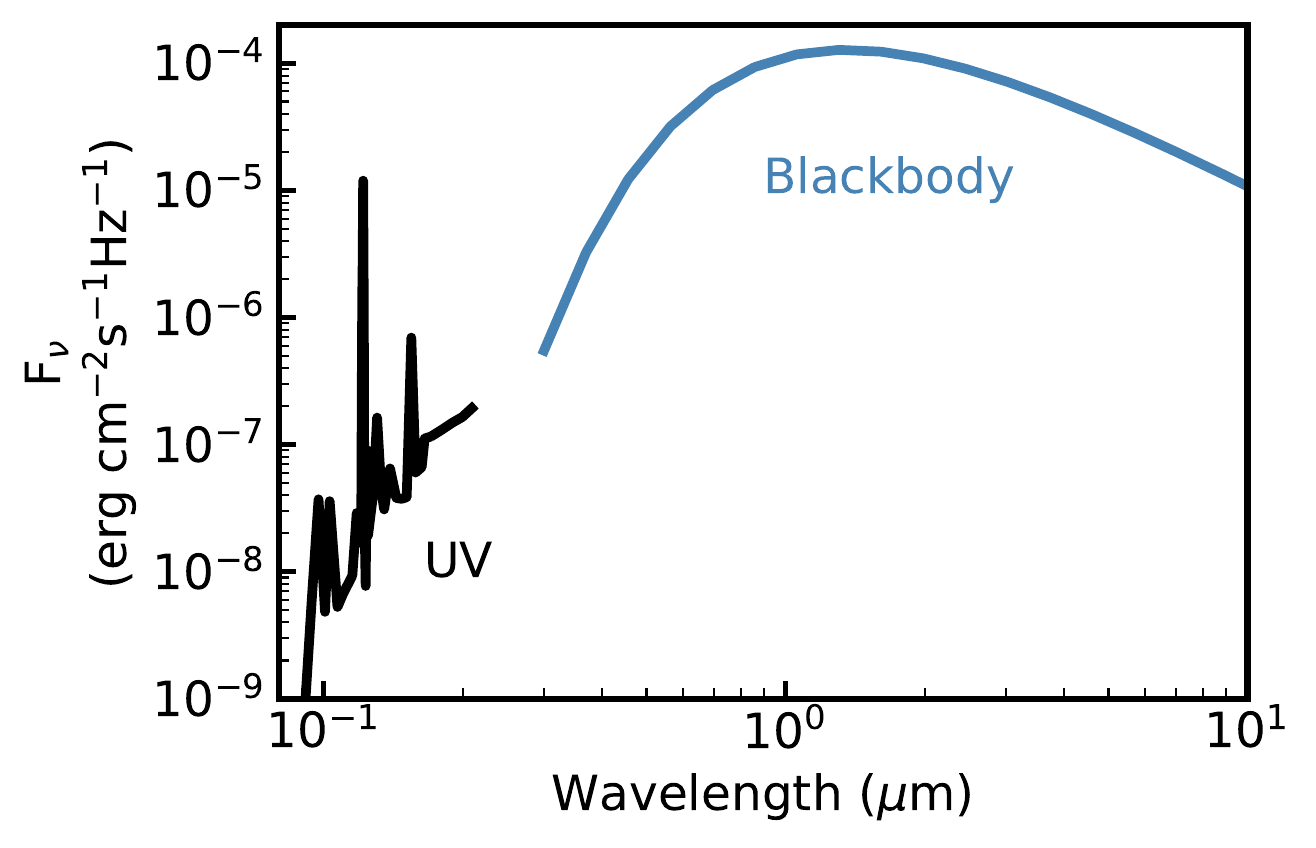}
	\caption{Example stellar spectrum showing the flux at the stellar surface, including blackbody and UV components for DM Tau ($T_{\mathrm{eff}}$ = 3350K, $L_{UV}$ = 4.0 $\times$10$^{-3}$ L$_\odot$).}
	\end{centering}
\label{fig:starspec}
\end{figure}

We use the radiative transfer code RADMC-3D \citep{Dullemond2012} {\fontfamily{qcr}\selectfont mctherm} task to solve for dust temperatures in the disk.  We assume a stellar spectrum consisting of a UV component, and a blackbody component corresponding to the star's effective temperature (4300K for the fiducial model).  We use the TW Hya UV spectrum shown in \citet{Cleeves2013} as a template for the UV component, scaled to reproduce the total UV luminosity of the star.  An example spectrum is shown in Figure \ref{fig:starspec}.  We also include an external UV field arising from the ISRF, where $G_0$ = 1 corresponds to a flux of 1.6 $\times$ 10$^{-3}$ erg cm$^{-2}$ s$^{-1}$ from 91 -- 200 nm \citep{Habing1968}.  Dust grain scattering and absorption cross-sections and anisotropy parameters are calculated using the DIANA project {\fontfamily{qcr}\selectfont Opacity-Tool} \citep{Woitke2016} for amorphous silicate:carbonaceous material.  The atmosphere dust population includes grain sizes from 5nm to 10$\mu$m and the midplane from 5nm to 1cm, in each case with a power law index of -3.5.  The thermal Monte Carlo calculation was performed in the anisotropic scattering mode with the Henyey-Greenstein formalism.  

We use the dust temperatures determined from the radiative transfer modeling to calculate gas temperatures, following the formalisms in \citet{Dartois2003} and \citet{Rosenfeld2013}.  In the midplane, the gas and dust are assumed to be thermally coupled such that $T_{gas}(r,z=0)$ = $T_m$ = $T_{dust}(r,z=0)$.  Above a height $z_q$ the gas temperature in the disk atmosphere $T_{gas}(r, z>z_q)$ = $T_a$ is described by:
\begin{equation}
T_a = T_{a,0} \Big{(}\frac{r}{R_{T,0}}\Big{)}^{-q}
\end{equation}
Our results are not very sensitive to the choice of gas temperature profile parameters, and to enable more direct comparisons we adopt a temperature gradient $q$ of 0.5 and a characteristic temperature $T_{a,0}$ of 150K for the T Tauri disks and 250K for the Herbig Ae disks at a radius $R_{T,0}$ = 10 AU. 

Between the midplane and an elevation $z_q$, the gas temperature is parameterized as:
\begin{equation}
T_{gas}(r,z) = T_{a} + (T_{m} - T_{a}) \Big{(}\mathrm{cos}\frac{\pi z}{2 z_q} \Big{)}^{2\delta}
\end{equation}
where $\delta$ is the shape of the vertical temperature gradient.  We fix $z_q$ to 2$H_{gas}$, where $H_{gas}$ is the gas scale height:
\begin{equation}
H_{gas} =  \frac{c_s}{\Omega} = \Big{(}\frac{kT_{gas, mid} r^3}{GM_{star}\mu m_H}\Big{)}^{1/2}.
\end{equation}
The gas density is then calculated by solving the hydrostatic equation, assuming a standard dust to gas ratio of 0.01.

To calculate carbon atom abundances, we assume that carbon is atomic and in the gas phase at disk heights where dissociative radiation can penetrate, which is set by attenuation of the UV field by dust, H and H$_2$, and CO.  We parameterize this depth as a vertical H$_2$ column density N$_{\mathrm{diss}}$ = 1.3 $\times$ 10$^{21}$ H$_2$ cm$^{-2}$ \citep{Qi2011, Rosenfeld2013, Williams2014}.  The photodissociation penetration depth will be impacted by factors including the UV intensity, grain growth, and carbon depletion \citep{Aikawa1999, Visser2009}, changing the depth at which atomic carbon is available in the disk.  We therefore include a penetration factor $f_{\mathrm{pen}}$, which effectively increases the H$_2$ column above which carbon atoms can exist ($N_{\mathrm{H_2}}$ = $f_{\mathrm{pen}}N_{\mathrm{diss}}$), to allow us to test how higher UV penetration impacts atomic carbon abundances in the disk.  To test the sensitivity of atomic carbon abundances to carbon depletion, we also include a depletion factor $f_{\mathrm{dep}}$.  Thus, the C atom abundance with respect to H$_2$ is:
\begin{equation}
x(C) = 
\begin{cases}
    1.4 \times 10^{-4} / f_{\mathrm{dep}} & \mathrm{if} \int_z^\infty n_{gas}(r, z')dz' \leq f_{\mathrm{pen}} N_{\mathrm{diss}} \\
    0              & \rm{else}
\end{cases}
\label{eq:xC}
\end{equation} 

To calculate the UV flux, we use the RADMC-3D task {\fontfamily{qcr}\selectfont mcmono} to sample the local radiation intensity $I_\nu$ in the disk model at wavelengths from 91 -- 200 nm and integrate:
\begin{equation}
F_{UV} = \int_{91nm}^{200nm}I_\nu d\nu.
\end{equation}

\noindent With these prescriptions for determining the local carbon atom abundances and UV fluxes in a disk, we can then test the effects of different disk physical properties.

\subsection{Sensitivity to disk physical properties}
\label{sec:grid_models}
\begin{figure}
	\includegraphics[width=\linewidth]{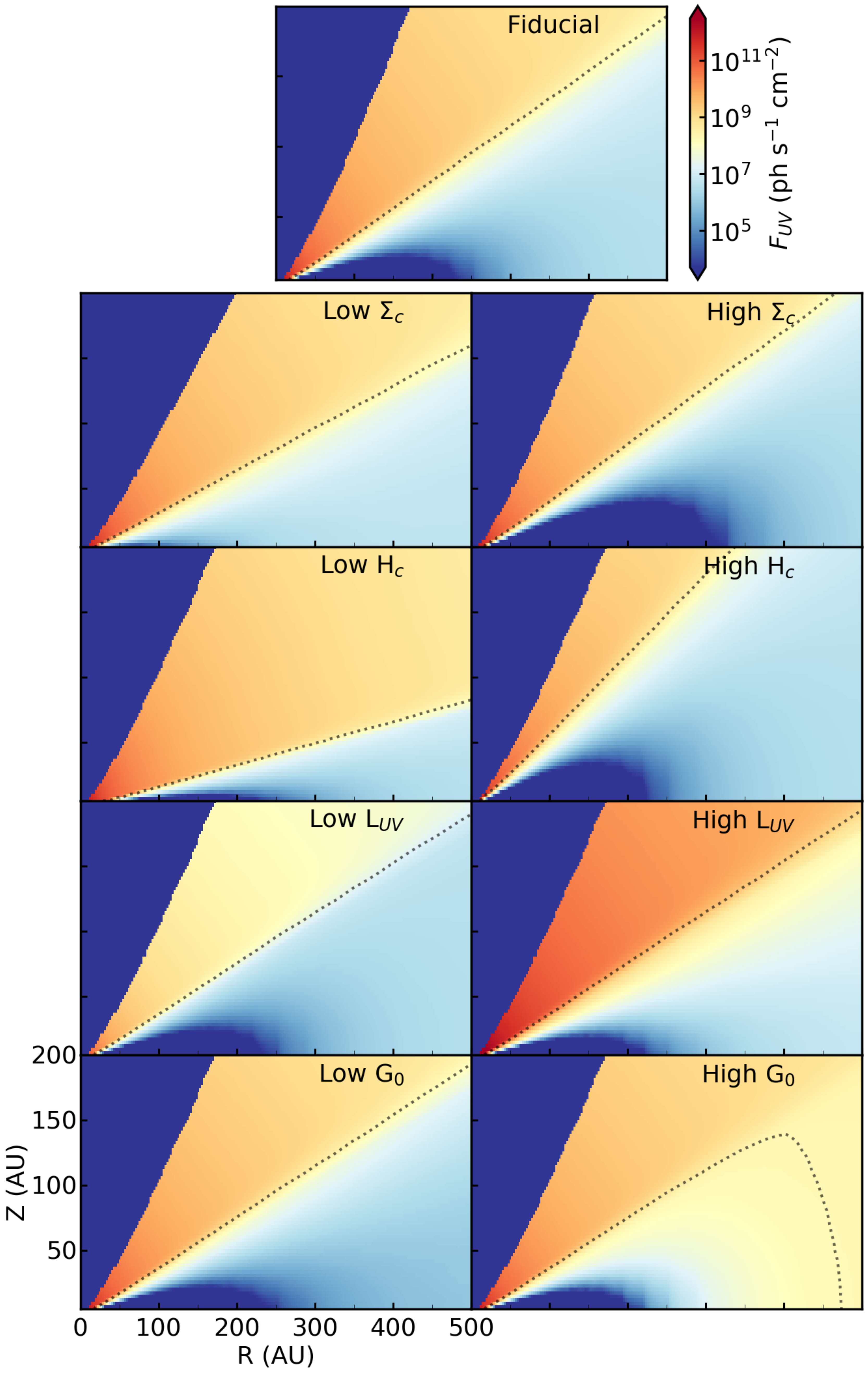}
	\caption{Modeling results exploring the effect of varying surface density ($\Sigma_c$), dust scale height ($H_c$), stellar UV luminosity ($L_{UV}$), and external UV field ($G_0$) on the local UV flux $F_{UV}$ in generic disk models.  The $\tau$ = 1 surface is shown for each disk as a dotted black line.}
\label{fig:UV_grid}
\end{figure}

Table \ref{tab:model_params} lists the fiducial model parameters as well as variations on the fiducial model.  Disk physical parameters are chosen to be within the range of source-specific values found in the literature (Section \ref{sec:disk_models}).  Figure \ref{fig:UV_grid} (top) presents the local UV fluxes resulting from our fiducial model.  

There is a clear drop in the UV flux below the  $\tau$ = 1 surface, which occurs around Z/R $\sim$0.4.  Another sharp drop in flux occurs in the inner $\sim$250 AU at low elevations, where high abundances of large dust grains result in large optical depths and negligible UV penetration.

\begin{deluxetable*}{lllllllllllll} 
	\tabletypesize{\footnotesize}
	\tablecaption{Parameters used in disk models \label{tab:model_params}}
	\tablecolumns{13} 
	\tablewidth{\textwidth} 
	\tablehead{
		\colhead{Model}           &
		\colhead{$R_{\mathrm{out}}$}       &
		\colhead{$R_{\mathrm{peb}}$}        &
		\colhead{$\Sigma_c$}              & 
		\colhead{$R_c$}              &
		\colhead{$\gamma$ }                        & 
		\colhead{$H_c$}                        & 
		\colhead{$R_H$}             &           
		\colhead{$h$}  & 
		\colhead{$T_{\star}$} &  
		\colhead{$L_{UV}$} & 
		\colhead{$R_{\star}$}  &
		\colhead{$G_0$}  \\
		\colhead{}           &
		\colhead{(AU)}       &
		\colhead{(AU)}        &
		\colhead{(g cm$^{-2}$)}              & 
		\colhead{(AU)}              &
		\colhead{}                        & 
		\colhead{(AU)}                        & 
		\colhead{(AU)}             &           
		\colhead{}  & 
		\colhead{(K)} & 
		\colhead{(L$_\odot$)} & 
		\colhead{(R$_{\odot}$)} &
		\colhead{} 
		}
\startdata
\multicolumn{13}{c}{Generic disk models} \\
\hline
Fiducial  & 600 & 300 & 0.5 & 100 & 0.8 & 10 & 100 & 1.2 & 4300 & 10$^{-3}$ & 1.6 & 5 \\
Low $\Sigma_c$ &  & & 0.05 & & & &  & &  & & &\\
High $\Sigma_c$ &  & & 2.5 & & & &  & &  & & & \\
Low $H_c$ &  & & & & & 4 &  & & &  & & \\
High $H_c$ &  & & & & & 16 &  & &  & & &  \\
Low $L_{UV}$ &  & & & & &  &  & &  &10$^{-4}$ & \\
High $L_{UV}$ &  & & & & &  &  & &  &10$^{-2}$ & \\
Low $G_0$ &  &  & & &  &  & & &  & & & 1 \\
High $G_0$ &  &  & & &  &  & & &  & & & 300 \\
\hline
\multicolumn{13}{c}{Source-specific disk models} \\
\hline
AS 209 & 400 & 205 & 0.04$^{[1]}$ & 126$^{[1]}$ & 0.4$^{[1]}$ & 13.3$^{[1]}$ & 100$^{[1]}$ & 1.1$^{[1]}$ & 4250$^{[2]}$ & 9.9 $\times$ 10$^{-5}$ $^{[3, a]}$ & 2.3 & 5 \\
MWC 480 & 600 & 175 & 0.48$^{[4]}$ & 81$^{[4]}$ & 0.75$^{[4]}$ & 16$^{[4]}$ & 100$^{[4]}$ & 1.25$^{[4]}$ & 8250$^{[5]}$ & 7.1 $\times$ 10$^{-2}$ $^{[6, b]}$ & 2.1 & 5 \\
IM Lup & 700 & 300 & 0.25$^{[7]}$ & 100$^{[7]}$ & 1.0$^{[7]}$ & 12$^{[7]}$ & 100$^{[7]}$ & 1.15$^{[7]}$ & 3850$^{[2]}$ & 9.0 $\times$ 10$^{-6}$ $^{[8, a]}$ & 2.9 & 5  \\
DM Tau$^c$ & 700 & 235 & 0.03$^{[9]}$ & 135$^{[9]}$ & 1.0$^{[9]}$ & 4.2$^{[9]}$ & 100$^{[9]}$ & 1.2$^{[9]}$ & 3350$^{[2]}$ & 4.0 $\times$ 10$^{-3}$$^{[10]}$ & 1.4 & 5  \\
HD 163296 & 800 & 165 & 0.07$^{[11]}$ & 150$^{[11]}$ & 0.8$^{[11]}$ & 16$^{[11]}$ & 150$^{[11]}$ & 1.35$^{[11]}$ & 9250$^{[5]}$ & 3.9 $\times$ 10$^{-2}$ $^{[6, b]}$ & 1.6 & 5  \\
LkCa 15$^c$ & 600 & 225 & 0.11$^{[9]}$ & 85$^{[9]}$ & 1.0$^{[9]}$ & 2.9$^{[9]}$ & 100$^{[9]}$ & 1.2$^{[9]}$ & 4250$^{[2]}$ & 3.7 $\times$ 10$^{-3}$ $^{[10]}$ & 1.7& 5 
\enddata
\tablenotetext{}{For the generic disk models, blank values indicate that the fiducial model value was adopted.} 
\tablenotetext{}{[1] \citet{Andrews2009}, [2] \citet{Gaia2016a, Gaia2016b, Gaia2018} [3] \citet{Andrews2007}, [4] \citet{Oberg2015}, [5] \citet{Vioque2018}, [6] \citet{Donehew2011}, [7] \citet{Cleeves2016}, [8] \citet{Gunther2010}, [9] \citet{Andrews2011}, [10] \citet{Yang2012}, [11] \citet{Rosenfeld2013a}}
\tablenotetext{}{(a) Calculated from the mass accretion rate via equations \ref{eq:luv_from_lacc} and \ref{eq:lacc_from_mdot}.\\(b) Calculated from the accretion luminosity via equation \ref{eq:luv_from_lacc} \\(c) Transition disks; dust densities are attenuated by a factor of 10$^{6}$ within 50 AU for LkCa 15 and 19 AU for DM Tau.}
\end{deluxetable*}

Figure \ref{fig:UV_grid} (bottom) shows the local UV fluxes resulting from variations on the fiducial model: low and high characteristic surface densities ($\Sigma_c$), dust scale heights ($H_c$), UV luminosities ($L_{UV}$), and external radiation fields ($G_0$).  Varying $\Sigma_c$ changes the total dust mass and therefore the optical depths in the disk, with the low and high $\Sigma_c$ cases showing enhanced and reduced UV penetration, respectively.  This is especially evident for the midplane but can also be seen in the location of the $\tau$ = 1 surface, which is pushed closer to the midplane for the low $\Sigma_c$ case.  Variations to the scale height $H_c$ result in large changes in the elevation of the $\tau$ = 1 surface, which is closer to the midplane in the low $H_c$ case.  This parameter also modestly impacts the location of the shielded midplane region, which is reduced in the low $H_c$ case.

The UV luminosity changes the magnitude of the UV fluxes in the disk models, with low and high $L_{UV}$ models showing lower and higher fluxes, respectively.  This effect is mainly important at high elevations in the inner disk.  The external UV field $G_0$ impacts the fluxes in deep disk layers in the outer disk, which are well shielded from stellar UV radiation. 

\begin{figure}
	\includegraphics[width=\linewidth]{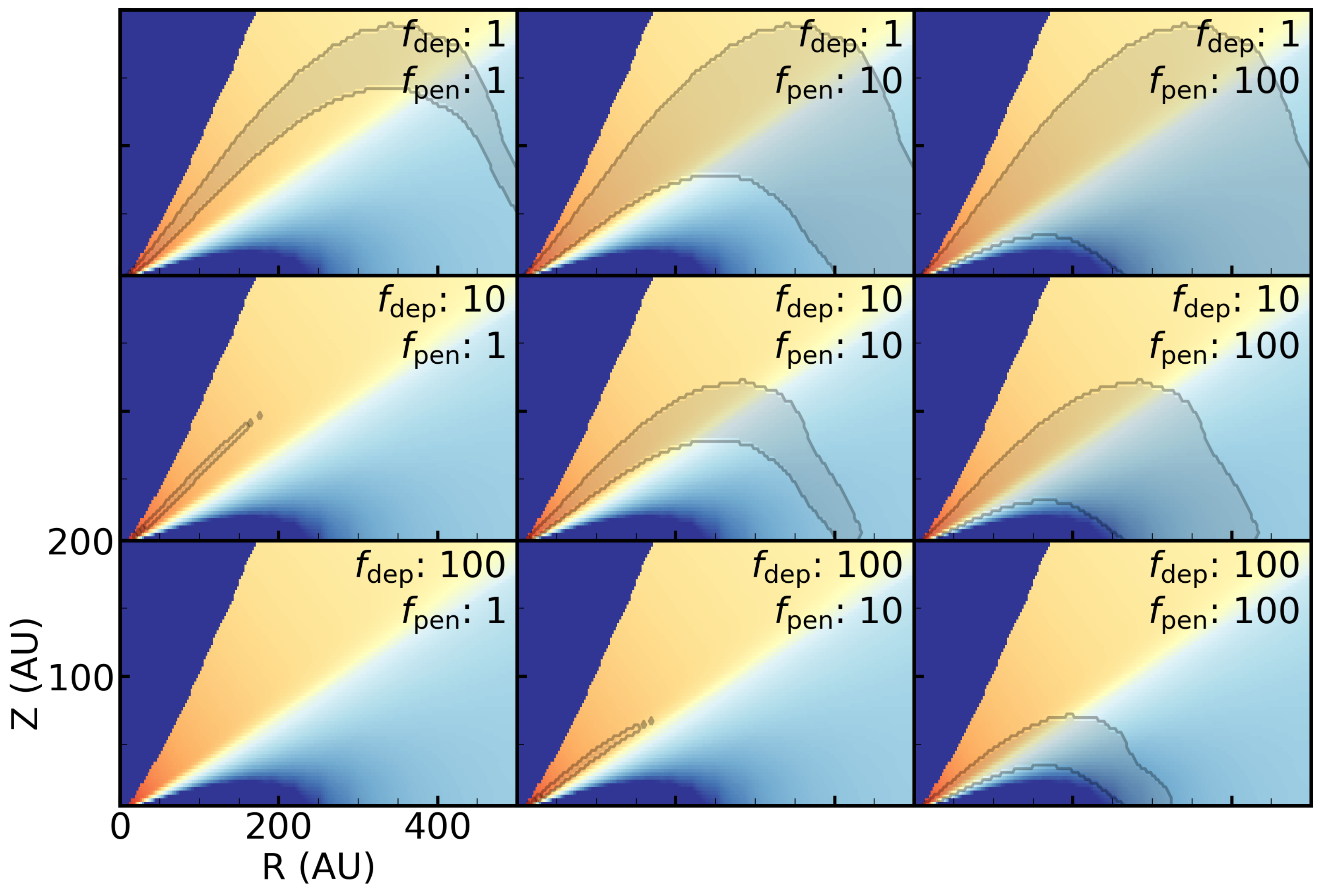}
	\caption{Modeling results exploring the effect of varying the UV pentration depth ($f_{\mathrm{pen}}$) and carbon depletion ($f_{\mathrm{dep}}$) on the atomic carbon availability in the fiducial disk model.  Shaded regions are bounded at low elevations by the C to CO transition, and at high elevations by where the atomic carbon abundance drops below 100 cm$^{-3}$.  UV fluxes are shown with the same color scale as in Figure \ref{fig:UV_grid}.}
\label{fig:C_grid}
\end{figure}

In addition to exploring the effects of the disk physical structure, we investigate how carbon depletion and photodissociation penetration impact the availability of atomic carbon in the disk.  Figure \ref{fig:C_grid} shows the disk regions where atomic carbon is abundant ($>$100 cm$^{-3}$) for a range of penetration factors $f_{\mathrm{pen}}$ and carbon depletion factors $f_{\mathrm{dep}}$, assuming the fiducial physical model.  The lower boundary of atomic carbon is set by the C to CO transition boundary, which is pushed to deeper layers in disks with higher values of $f_{\mathrm{pen}}$.  The upper boundary of atomic carbon is set as the location where the carbon abundance drops below 100 cm$^{-3}$, which occurs at lower elevations for more carbon-depleted systems.  This value is somewhat arbitrary, but was chosen to demonstrate the impact of carbon depletion on atomic carbon availability.

When the magnitude of $f_{\mathrm{pen}}$ and $f_{\mathrm{dep}}$ are matched, atomic carbon is abundant in a thin strip of elevations in the disk.  This strip occurs deeper in the disk for higher penetration and carbon depletion factors, as both the CO photodissociation boundary and the C abundance boundary are pushed to lower elevations.  When UV penetration is increased and there is no carbon depletion, the strip of atomic carbon is much wider due to the lower elevation of the CO transition boundary.  Similarly, when carbon is depleted and there is no increase in UV penetration, the zone of abundant atomic carbon is negligible.  

While we vary $f_{\mathrm{pen}}$ and $f_{\mathrm{dep}}$ independently in these toy models, in reality they are linked: carbon depletion will decrease CO self-shielding, resulting in a higher $f_{\mathrm{pen}}$.  Because of this, scenarios with $f_{\mathrm{dep}}$ $>$ $f_{\mathrm{pen}}$ are not physically likely. 

Having established that moderate changes in the disk physical and chemical parameters may strongly impact the distributions of high-flux and atomic carbon zones, we now explore these effects in a set of source-specific models to enable a comparison with observed C$_2$H morphologies.

\begin{figure*}
	\includegraphics[width=\linewidth]{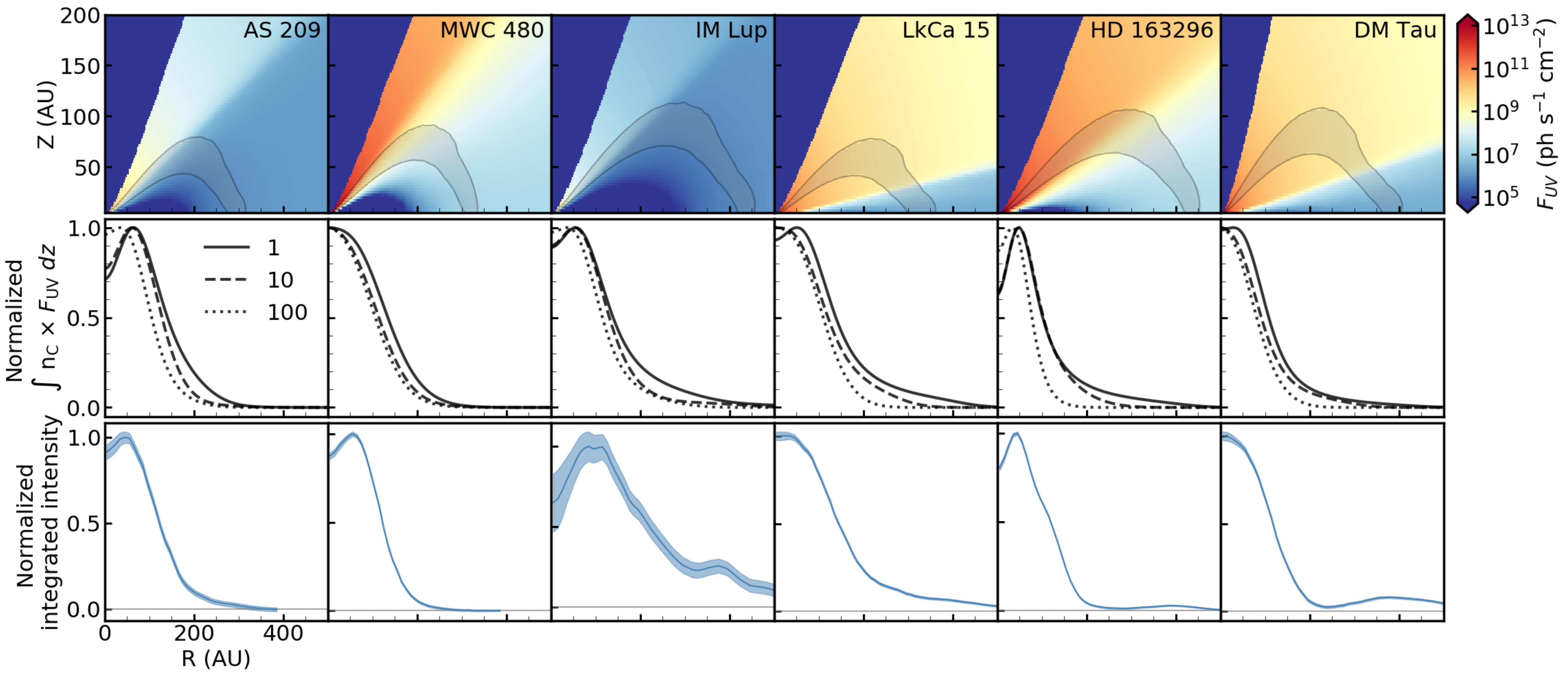}
	\caption{Top: modeled local UV fluxes (color map) and atomic carbon (shaded grey) in a subset of sources from this survey, shown for $f_{dep} = f_{pen} = 10$.  Middle: overlap profiles showing the vertically-integrated convolution of UV fluxes and carbon atom abundance, for  $f_{dep} = f_{pen}$ = 1, 10, and 100.  Profiles have been smoothed with the C$_2$H restoring beam size.  Bottom: C$_2$H J = $\frac{7}{2}$ -- $\frac{5}{2}$ deprojected and radially averaged emission profiles, reproduced from Figure \ref{fig:radprof}.}
\label{fig:UV_C_sources}
\end{figure*}

\subsection{C$_2$H morphologies in our disk sample}
\label{sec:disk_models}
We explore the relationship between disk physical properties and C$_2$H morphology by modeling a subset of the disk sample that covers the full range in observed morphologies: AS 209, MWC 480, IM Lup, DM Tau, HD 163296, and LkCa 15.  The physical structure for each of these disks has been modeled previously in the literature, and we use these constraints to produce toy models for each disk (Table \ref{tab:model_params}).  To convert from literature disk masses to $\Sigma_c$ we use the convention described in \citet{Andrews2009}:
\begin{equation}
\Sigma_c = \frac{(2 - \gamma) M_{\mathrm{disk}}}{2\pi R_c^2}
\end{equation}

When possible we use measured UV luminosities $L_{UV}$ to scale the UV spectrum shown in Figure \ref{fig:starspec} for each host star.  When the UV luminosity has not been measured for a star, we convert from the accretion luminosity $L_{\mathrm{acc}}$using the relation derived in \citet{Yang2012}:
\begin{equation}
\mathrm{log}_{10}(L_{UV}) = -1.670 + 0.836\mathrm{log}_{10}(L_{\mathrm{acc}}),
\label{eq:luv_from_lacc}
\end{equation}
and when an accretion luminosity is not available we estimate it from the mass accretion rate following \citet{Gullbring1998}:
\begin{equation}
L_{\mathrm{acc}} = 0.8 \times G \dot{M}_\star/R_\star.
\label{eq:lacc_from_mdot}
\end{equation}

Outer disk radii $R_{\mathrm{out}}$ and pebble disk radii $R_{\mathrm{peb}}$ are estimated from the extent of C$^{18}$O and the millimeter continuum respectively (Figure \ref{fig:radprof}).

$G_0$ for each disk depends on the population of nearby massive stars. \citet{Cleeves2016} find a low value of $G_0$ $\leq$ 4 for IM Lup, consistent with its location in a low-mass star-forming region.  Because $G_0$ is not constrained for the other disks in our sample, we adopt a conservative uniform value of 5, with the caveat that disks with more high-mass stellar neighbors will experience a higher $G_0$ leading to more UV penetration in the outer disk compared to our models.

For all disks, we evaluate atomic carbon abundances for $f_{\mathrm{pen}}$ = $f_{\mathrm{dep}}$ = 1, 10, and 100, taking into account that these two parameters should generally be closely related.  
Figure \ref{fig:UV_C_sources} (top panels) shows the atomic carbon and UV flux distributions for the $f_{dep} = f_{pen} = 10$ case for the 6 modeled disks.  Since we expect C$_2$H formation to be enhanced in regions with overlapping atomic carbon and high UV fluxes, we also plot the vertically-integrated convolution of both factors: $\int n_c \times F_{UV} dz$, hereafter referred to as overlap profiles.  Lastly, to enable a comparison between the overlap profiles and our observations, we reproduce the C$_2$H N = 3 -- 2, J = $\frac{7}{2}$ -- $\frac{5}{2}$ radial profiles from Figure \ref{fig:radprof}.

The shapes of the modeled overlap profiles and the observed C$_2$H emission profiles generally match well.  The key features of the C$_2$H emission profiles that we are aiming to reproduce are central peaks vs. central depressions, compact vs. broad extents, and the presence of outer rings.

LkCa 15 and DM Tau show centrally peaked C$_2$H emission, which is reproduced for overlap profiles with moderate to high depletion ($f_{dep}$ = 10--100) in these disks.  The remaining disks show centrally depressed emission, which is reproduced for all disks except MWC 480.  For AS 209, IM Lup, and HD 163296, a central depression is present for all depletion factors, though higher depletions appear to be in better agreement with the observed C$_2$H profiles in AS 209 and HD 163296.

AS 209 and MWC 480 both exhibit compact C$_2$H emission with no extended features, and their overlap profiles are similarly compact.  Regardless of the depletion level, the physical structures of these disks do not facilitate C$_2$H production at large radii.  IM Lup and LkCa 15 both show extended C$_2$H emission at large radii ($>$200 AU), which is reproduced by the low-depletion ($f_{dep}$ = 1--10) overlap profiles.  HD 163296 and DM Tau show compact inner components as well as rings external to the dust continuum.  This shape can be reproduced by high depletion (10--100) in the inner 100--200 AU and low depletion beyond this.

In summary, higher depletion factors typically result in more pronounced central depressions and more compact emission, while lower depletion factors lead to more extended profiles.  Across the disk sample, the observed emission is best described by high depletions (10--100) in the inner 100--200 AU, and low depletions (1--10) at larger disk radii.  Still, while the shape of each overlap profile is somewhat dependent on the depletion level, it is more sensitive to the disk physical structure: the variation in overlap profiles across the disk sample is greater than the effects of depletion on any given disk profile.  While more rigorous chemical modeling is needed to explore the origin of these varied emission features in more detail, the interplay between local UV fluxes and atomic carbon availability is a useful framework that can explain the wide range in observed C$_2$H emission morphologies in our sample.

\section{Discussion}
\label{sec:disc}
\subsection{C$_2$H and HCN emission origin}
\label{sec:emission_origin}
\begin{figure}
	\includegraphics[width=\linewidth]{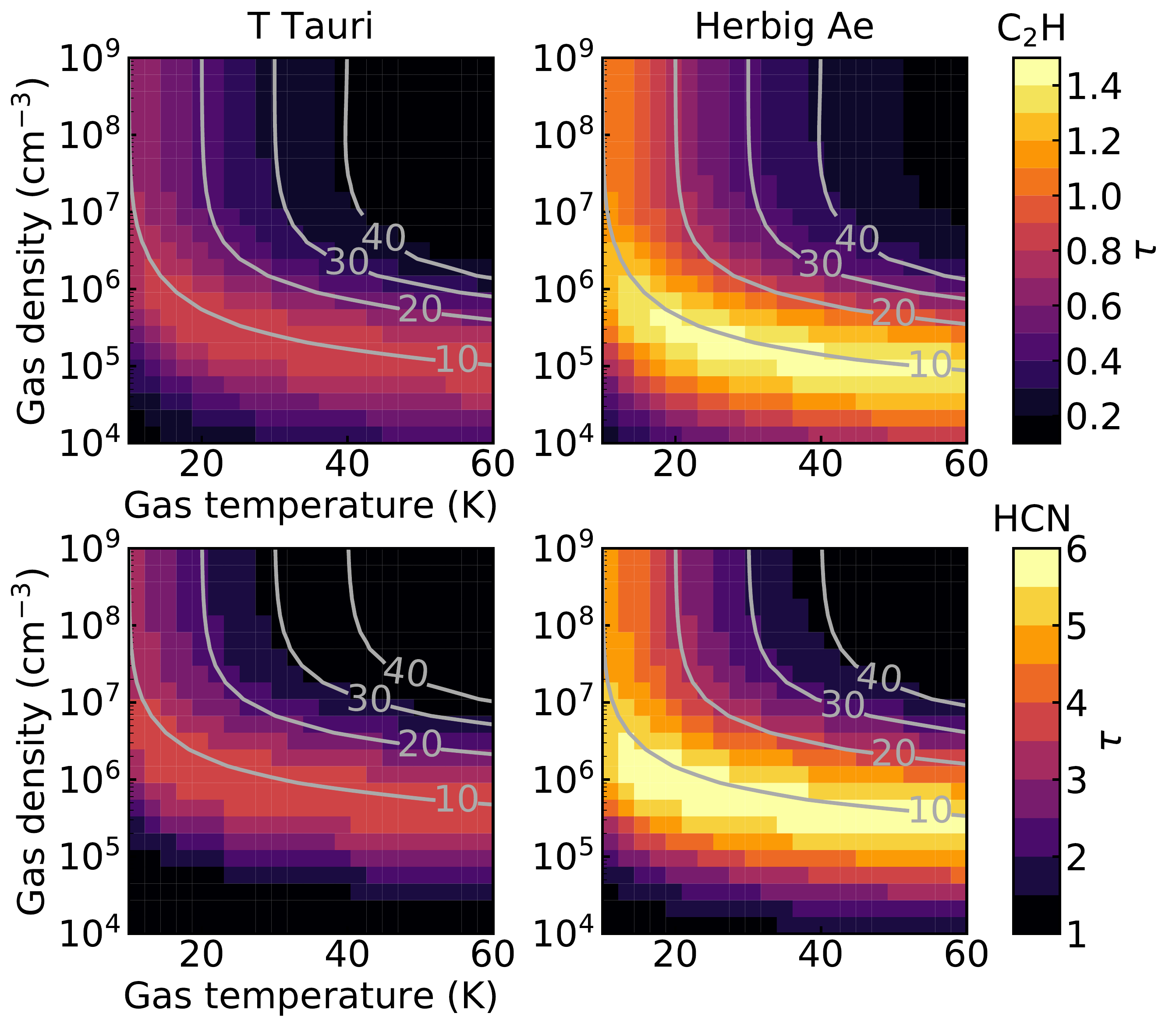}
	\caption{Radex predictions of the C$_2$H (top) and HCN (bottom) optical depth (color map) and excitation temperature (grey lines) resulting from a range of gas densities and temperatures.  Left and right panels show results given column densities and line widths for a generic T Tauri and Herbig Ae disk, respectively.  Optical depths are for the main hyperfine component (see Table \ref{tab:linedat}).}
\label{fig:radex}
\end{figure}

From hyperfine fitting and isotopologue analysis of C$_2$H and HCN, we find typical rotational temperatures of 10--40K and 5--20K and optical depths of 0.5 -- 5 and 3 -- 10, respectively.  In order to explore the excitation conditions further, we run Radex models \citep{Vandertak2007} for a grid of densities and temperatures typical within a disk.  For each molecule we model generic T Tauri and Herbig Ae conditions.  The C$_2$H T Tauri and Herbig Ae models assume column densities of 5 $\times$10$^{13}$ cm$^{-2}$ and 2 $\times$ 10$^{14}$ cm$^{-2}$ and line widths of 0.2 km s$^{-1}$ and 0.5 km s$^{-1}$ respectively; the HCN T Tauri and Herbig Ae models use column densities of 1 $\times$10$^{13}$ cm$^{-2}$ and 3 $\times$ 10$^{13}$ cm$^{-2}$ and line widths of 0.4 km s$^{-1}$ and 0.8 km s$^{-1}$ respectively.  These choices are informed by the column densities and line widths derived from our observations.  The resulting optical depths and excitation temperatures are shown in Figure \ref{fig:radex}.

We are able to reproduce very cold (10K) and moderately optically thick C$_2$H and HCN emission assuming either very cold and dense conditions, or warmer and less dense conditions.  These scenarios correspond to either thermal emission from the disk midplane, or sub-thermal emission from the disk atmosphere.  Our disk physical models (Figure \ref{fig:UV_C_sources}) suggest that C$_2$H formation is favored at moderate disk elevations (Z/R $\sim$0.4), which would correspond to sub-thermal emission from warm gas.  Indeed, full disk chemistry models similarly predict abundant C$_2$H at higher disk elevations \citep{Bergin2016, Cleeves2018}).  Likewise, based on the flux ratio of C$_2$H N = 4 -- 3/N = 3 --2 in TW Hya, \citet{Kastner2015} and \citet{Bergin2016} conclude that C$_2$H emission arises from warm gas temperatures (40--50K).

Another possible origin of the low excitation temperatures that we derive is beam dilution due to chemical substructure.  High-resolution continuum observations are revealing that dust substructure is widespread in disks; indeed, four disks in our sample, AS 209, HD 163296, HD 143006, and IM Lup, were observed as part of the DSHARP program \citep{Andrews2018}, and each exhibits substructure (spiral arms or wide gaps) at high resolution.  If molecular line emission is similarly structured, this would result in substantial beam dilution at the resolution of our observations.  High-resolution molecular line observations are required to determine if gas substructure is prevalent, and if so to obtain more accurate constraints on the molecular excitation temperatures and column densities.

\subsection{Disk-averaged C$_2$H and HCN column densities}
Assuming similar excitation conditions, the disk-averaged C$_2$H and HCN column densities across the source sample are quite consistent (Table \ref{tab:Nts}).  For any given excitation temperature and optical depth, C$_2$H column densities vary by a factor of $\sim$7 and HCN column densities by a factor of $\sim$10.  This indicates that the chemistry of these small molecules in disk atmospheres is robust even across the wide range of physical environments probed in our sample.  Modeling is required to determine if this translates to consistent midplane abundances of these molecules as well.  In particular, given the importance of HCN for origins of life chemistry on Earth \citep{Patel2015, Sutherland2016}, determining if HCN is reliably available in planet-forming disk midplanes is a key consideration in evaluating the potential for prebiotic chemistry in other solar systems.  The radially resolved excitation temperatures and column densities derived in this work should enable benchmarking of source-specific chemical models that can address this question.

\subsection{C$_2$H column density profiles: tracing planet locations?}
\begin{figure}
	\includegraphics[width=0.9\linewidth]{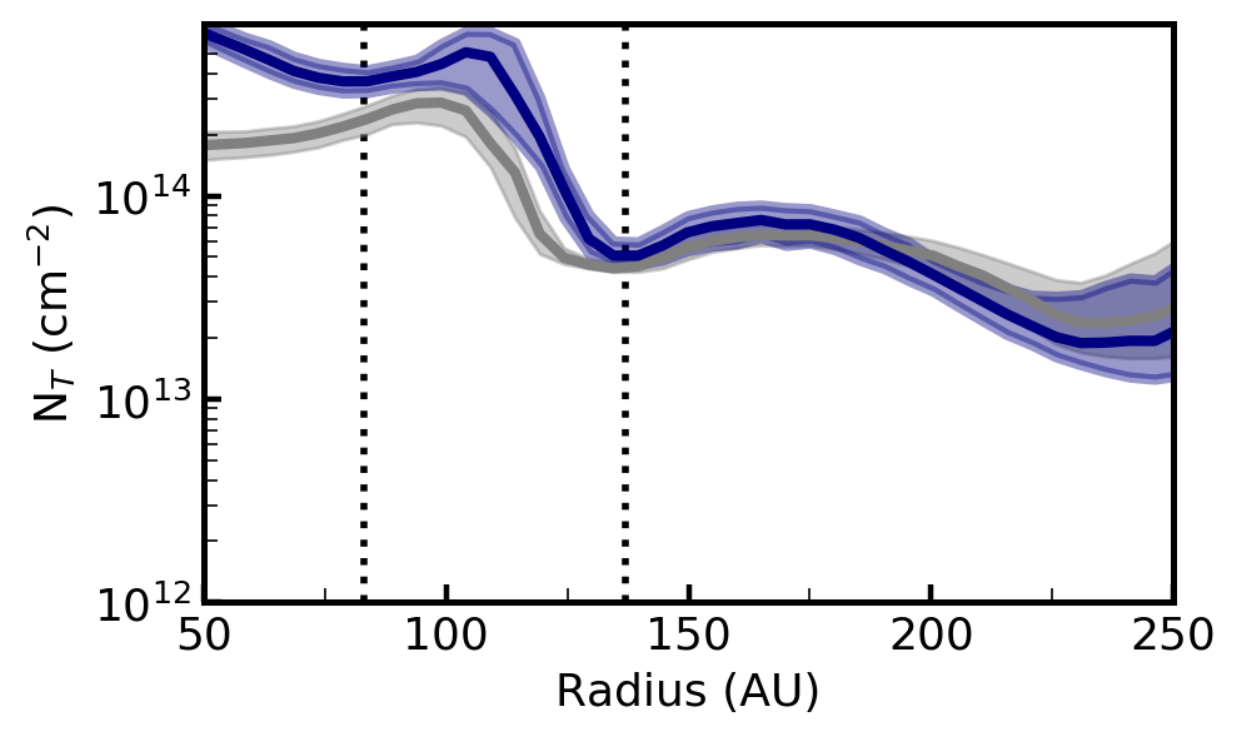}
	\caption{HD 163296 column density profile from Figure \ref{fig:taurad_c2h}, overplotted with the locations of Jupiter-mass planets at 83 AU and 137 AU identified by \citet{Teague2018}.}
\label{fig:hd163296}
\end{figure}

The C$_2$H column density profile for HD 163296 shows particularly sharp features compared to the other disks.  Interestingly, the locations of local minima in the column density profile coincide nearly exactly with the locations of planets identified by \citet{Teague2018} (Figure \ref{fig:hd163296}), raising the possibility that well-resolved C$_2$H column density profiles could potentially trace the locations of large planets.  Indeed, this is not surprising if planets coincide with perturbations in the gas density profile, and if the C$_2$H column density traces the overall gas density in the disk.  We note that this effect should not be limited to C$_2$H, but that the C$_2$H is a good tracer because its hyperfine structure enables a robust determination of the column density profile.  This method could potentially be a valuable tool to find or confirm the presence of planets in disks.  It is likely that this method is sensitive only to massive planets given that the gas surface density would need to be significantly perturbed to be detectable in the C$_2$H column density.  HD 163296 is the only disk in our sample with well-detected C$_2$H substructure.  C$_2$H observations with high sensitivity and spatial resolution in other disks with planet candidates are required to test whether precise column density measurements derived from hyperfine fitting can be used as tracers of planets in disks.

\subsection{Diversity in C$_2$H morphologies}
In Section \ref{sec:diskmodels} we explored the origin of the observed C$_2$H morphologies by producing models of the local UV fluxes and atomic carbon abundances in a subset of the disks targeted in this survey.  The overlap between these factors reproduces the shapes of the observed C$_2$H radial emission profiles, assuming moderate carbon depletion ($f_{dep}$ = 10--100) in the inner 100 -- 200 AU and low carbon depletion ($f_{dep}$ = 1--10) in the outer disk.  The enhancement of gas-phase carbon in the outer disk could be due to UV photodesorption of CO ice at large radii, where a decreased dust opacity allows for greater UV penetration.  This effect is thought to be responsible for the outer DCO$^+$ ring in IM Lup \citep{Oberg2015b} and would naturally explain why our models indicate that carbon abundances are enhanced beyond the dust continuum.  Other factors may also contribute to carbon enhancements beyond $\sim$200 AU, including slower chemistry resulting in less conversion of CO into less volatile carbon carriers, and reduced grain settling at large radii.

Previous modeling of C$_2$H in disks has explored the effect of the gas-phase C/O ratio as a driver of compact vs. ringed morphologies \citep{Bergin2016, Cleeves2018}.  Here, without explicitly considering oxygen, we have demonstrated that a wide range in C$_2$H morphologies (centrally peaked and centrally depressed; compact and extended; ringed and ringless) can be explained based on the UV field, disk structure, and carbon abundance alone.  A high C/O ratio may help to achieve the high atomic carbon abundances required for C$_2$H formation, but our results underscore that additional physical drivers must be considered in addition to this chemical driver.

\subsection{Chemical relations between C$_2$H, HCN, and C$^{18}$O}
The chemical relationship between C$_2$H, HCN, and C$^{18}$O in our disk sample should provide constraints on the elemental (C, N, O) composition in elevated disk layers, a key target of many disk chemistry studies.  In the absence of reliable abundance constraints, and with column density constraints towards only a subset of our sources, we instead explore correlations between normalized line fluxes.  We normalize each disk-integrated line flux to the continuum flux in order to cancel out any dependency on the disk angular size.  

\begin{figure*}
	\centering
	\includegraphics[width=0.9\linewidth]{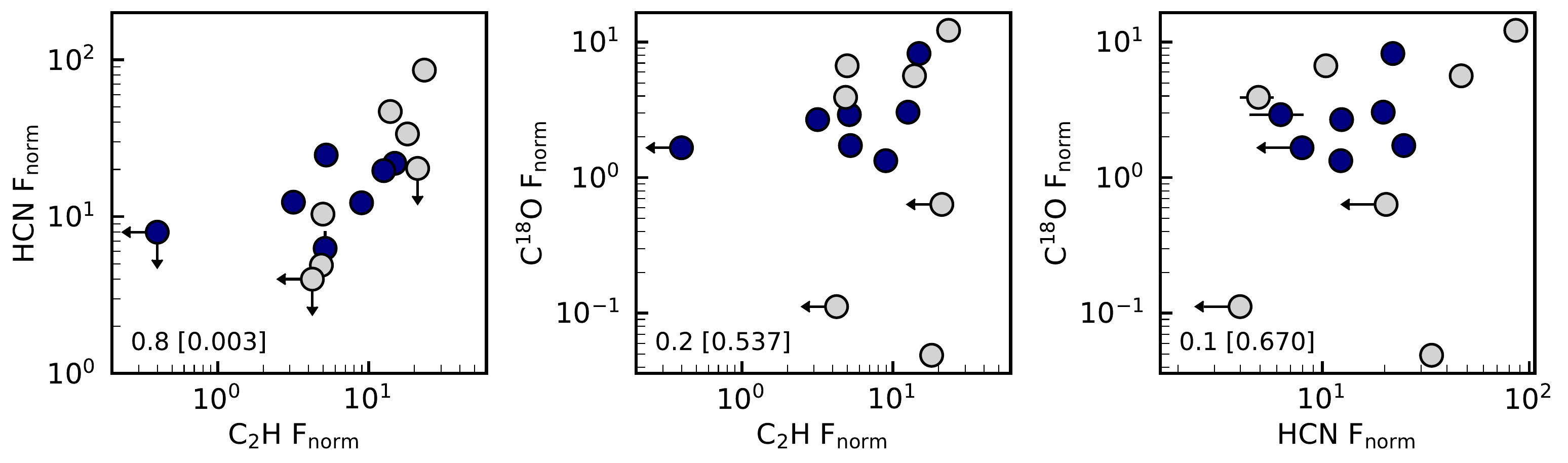}
	\caption{C$_2$H, HCN, and C$^{18}$O flux correlation plots.  Each disk-integrated line flux is normalized to the disk-integrated continuum flux.  Blue markers indicate young ($<$5Myr) disks and grey markers indicate old ($>$5Myr) disks.  Spearman correlation coefficients are indicated in the bottom left corner, with corresponding p-values in brackets.}
\label{fig:ratio_cno}
\end{figure*}

Figure \ref{fig:ratio_cno} shows the normalized fluxes of each species plotted against one another.  C$_2$H and HCN fluxes show a positive correlation, but there is no clear relationship between C$^{18}$O and C$_2$H or between C$^{18}$O and HCN.  This is reflected in the Spearman correlation coefficients (excluding upper limits): for C$_2$H with HCN the coefficient is 0.8 with a p-value of 0.003, i.e. a statistically significant trend.  For C$_2$H with C$^{18}$O and HCN with C$^{18}$O, the correlation coefficients are close to zero (0.2 and 0.1 respectively), indicating no trend.  Both these values are statistically insignificant, and it is possible that probing more disks with a wider range of normalized fluxes will reveal some trend.

The positive correlation between HCN and C$_2$H suggests that these molecules share a common physical or chemical driver.  This is consistent with a scenario in which cyanide and hydrocarbon formation are accelerated by oxygen removal from the gas phase via H$_2$O and CO depletion \citep[e.g.][]{Du2015, Bergin2016}.  However, we observe no clear relationship between C$_2$H or HCN with C$^{18}$O, suggesting that either (i) C$^{18}$O is not a good tracer of the oxygen abundance in the disk atmosphere or (ii) another mechanism is driving C$_2$H and HCN formation.  Moreover, in a framework where volatiles are sequestered in the midplane via grain growth and settling, it is expected that disks will become more oxygen depleted as they age, which should drive HCN and C$_2$H formation in older disks.  In our sample, however, there is no clear relation between older disks and higher C$_2$H and HCN fluxes (Figure \ref{fig:ratio_cno}), suggesting that age is not the main driver of this trend.

While these observations are in some conflict with an oxygen depletion mechanism, an important caveat to this analysis is that fluxes may not be reliable tracers of molecular abundances.  We have demonstrated in Section \ref{sec:tau_cd} that at least C$_2$H and HCN are often optically thick in our sample.  Moreover, excitation effects will change the relationship between fluxes and abundances in different physical environments.  We emphasize the need for deeper follow-up observations of optically thin isotopologues of HCN and C$_2$H to properly explore correlations between molecular column densities across the entire disk sample.  Ultimately, column density constraints for all three molecules and H$_2$ would enable an evaluation of abundance correlations, which will be required to firmly confirm or reject an oxygen depletion mechanism.  In any event, there are likely other factors at play that impact the chemistry of these small molecules, and more work remains to understand the mechanisms and outcomes of oxygen depletion in disk environments with diverse physical properties.

\section{Conclusions}
We have completed a survey of the small molecules C$_2$H, HCN, and C$^{18}$O in 14 protoplanetary disks using ALMA and the SMA.  When possible, we extract optical depth, excitation temperature, and column density radial profiles for C$_2$H and HCN using hyperfine fitting and isotopologue ratios.  To explore relationships between disk physical properties and the observed C$_2$H morphologies, we produce toy models of the local UV fluxes and carbon atom abundances in a grid of physical models as well as source-specific models.  Based on this analysis, we conclude the following:
\begin{enumerate}
\item C$_2$H emission is marginally optically thick in the inner $\sim$200AU of most disks, with optical depths around 1--5.  HCN emission is optically thicker, with optical depths around 3--10. 

\item Excitation temperatures range from 10 -- 40K for C$_2$H, and 5 -- 20K for HCN.  These low temperatures likely result from sub-thermal emission in the warm disk atmosphere, and/or beam dilution due to chemical substructure.

\item Assuming similar excitation conditions, C$_2$H column densities vary by a factor of $\sim$7 and HCN by a factor of $\sim$10 across the disk sample.  This indicates that the abundances of these small molecules in disk atmospheres are fairly insensitive to physical properties of the disk.

\item The HD 163296 C$_2$H column density profile is well-resolved and shows local minima coinciding with the radii where giant planets have been identified by \citet{Teague2018}.  This raises the prospect that well-constrained C$_2$H column density profiles may trace planet locations if they cause large perturbations to the total gas surface density profile.

\item The wide range of C$_2$H emission morphologies can be explained by the interplay between the local UV fluxes and atomic carbon abundances throughout the disk.  Disk-dependent overlap between atomic carbon zones and a high UV flux gives rise to compact vs. extended and ringed vs. ringless C$_2$H morphologies.  Our models indicate moderate carbon depletion in the inner 100-200 AU and minimal carbon depletion at larger radii.

\item We see a correlation between HCN and C$_2$H fluxes, indicating that these species share a common physical or chemical driver.  The lack of (i) correlations with C$^{18}$O and (ii) a clear evolutionary trend to C$_2$H and HCN enhancement throw into question whether gas-phase oxygen depletion is a driver of hydrocarbon and cyanide chemistry.   More observations and theoretical work are needed to elucidate the relationships between C, N, and O chemistries in disks.
\end{enumerate}

\acknowledgments 
This paper makes use of ALMA data, project codes: 2013.1.00226, 2015.1.00671.S, 2015.1.00964.S, and 2016.1.00627.S. ALMA is a partnership of ESO (representing its member states), NSF (USA), and NINS (Japan), together with NRC (Canada) and NSC and ASIAA (Taiwan), in cooperation with the Republic of Chile. The Joint ALMA Observatory is operated by ESO, AUI/NRAO, and NAOJ. The National Radio Astronomy Observatory is a facility of the National Science Foundation operated under cooperative agreement by Associated Universities, Inc.  

The Submillimeter Array is a joint project between the Smithsonian Astrophysical Observatory and the Academia Sinica Institute of Astronomy and Astrophysics and is funded by the Smithsonian Institution and the Academia Sinica.

The authors are grateful to L. Cleeves and L. Kelley for valuable feedback. J.B.B acknowledges funding from the National Science Foundation Graduate Research Fellowship under Grant DGE1144152.   This work was supported by an award from the Simons Foundation (SCOL \# 321183, KO).

\software{
{\fontfamily{qcr}\selectfont NumPy} \citep{VanDerWalt2011},
{\fontfamily{qcr}\selectfont Matplotlib} \citep{Hunter2007},
{\fontfamily{qcr}\selectfont Astropy} \citep{Astropy2013}, 
{\fontfamily{qcr}\selectfont emcee} \citep{Foreman-Mackey2013},
{\fontfamily{qcr}\selectfont RADMC-3D} \citep{Dullemond2012}
}

\FloatBarrier
\clearpage

\appendix 
\section{Observational details}
\label{sec:app_obsdat}
\FloatBarrier
Tables \ref{tab:alma_details} and \ref{tab:sma_details} list the ALMA and SMA observation details relevant to this project.  Similar details for the C$^{18}$O and CO observations can be found in J. Pegues (in preparation) and \citet{Huang2017}.  

\begin{figure}
	\includegraphics[width=\linewidth]{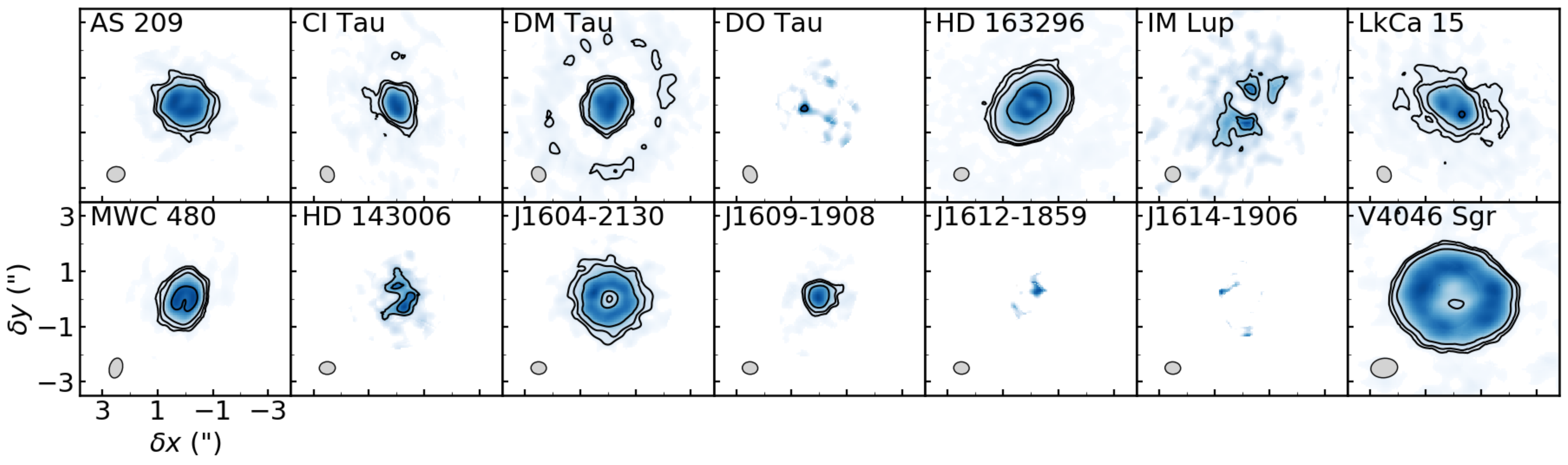}
	\caption{Moment zero maps for C$_2$H N = 3 -- 2, J = $\frac{5}{2}$ -- $\frac{3}{2}$, F = 3 -- 2 and 2 -- 1 transitions, extracted with Keplerain masking.  Contour levels show 3, 5, 10, 30, and 100 $\times$ the median rms.  Restoring beams are shown in the lower left of each panel.}
\label{fig:mom0_c2hj5}
\end{figure}

\begin{figure}
\begin{centering}
	\includegraphics[width=0.95\linewidth]{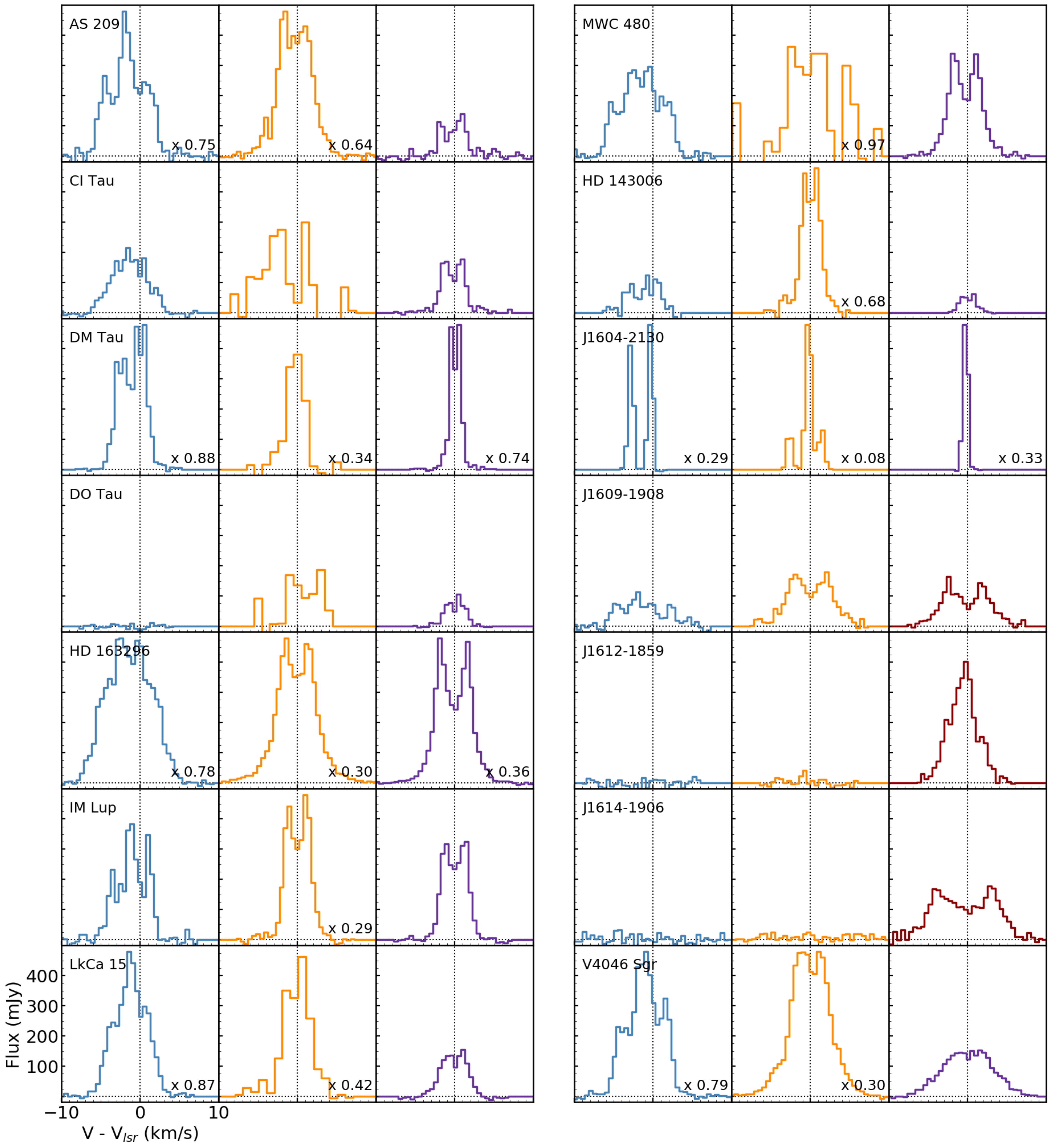}
	\caption{Disk-integrated spectra extracted with Keplerian masks for the C$_2$H J = $\frac{7}{2}$ -- $\frac{5}{2}$ (left), HCN J = 3 -- 2 (middle), and C$^{18}$O or $^{12}$CO J = 2 -- 1 (right) transitions.  Spectra have been scaled by the indicated amount for clarity.}
\end{centering}
\label{fig:spec}
\end{figure}

Table \ref{tab:fluxes_other} lists the line observation details for the H$^{13}$CN J = 3 -- 2, C$^{18}$O and CO J = 2 -- 1, and C$_2$H N = 3 -- 2,  J = $\frac{5}{2}$ -- $\frac{3}{2}$ transitions.  Figure \ref{fig:mom0_c2hj5} shows moment zero maps for the C$_2$H J = $\frac{5}{2}$ -- $\frac{3}{2}$ transition.  Figure \ref{fig:spec} shows disk-integrated spectra for each molecule.

\begin{deluxetable}{lccccccc} 
	\tabletypesize{\footnotesize}
	\tablecaption{C$_2$H and HCN ALMA observations \label{tab:alma_details}}
	\tablecolumns{8} 
	\tablewidth{\textwidth} 
	\tablehead{
		\colhead{Source}           &
		\colhead{\# Ant.}        &
		\colhead{Baselines (m)}              & 
		\colhead{Total on-source }              &
		\colhead{Date}       &
		\colhead{Bandpass cal.}                        & 
		\colhead{Phase cal.}                        & 
		\colhead{Flux cal.}                        \\
		\colhead{} & 
		\colhead{} & 
		\colhead{ (m)} & 
		\colhead{ time (min)} &
		\colhead{}  & 
		\colhead{} & 
		\colhead{} &
		\colhead{}
		}
\startdata
\multicolumn{8}{c}{Cycle 3 -- 2015.00964.S} \\
\hline
HD 143006 &  48 & 15 -- 704 & 14 & 2016 05 17 & J1517-2422 & J1551-1755 & Titan \\
 &  &  & & 2016 05 23 & J1517-2422 & J1551-1755 & Titan \\
 &  &  & & 2016 06 12 & J1517-2422 & J1551-1755 & J1517-2422 \\
J16042165-2130284 & 48 & 15 -- 704 & 21 & 2016 05 17 & J1517-2422 & J1551-1755 & Titan \\
 &  &  & & 2016 05 23 & J1517-2422 & J1551-1755 & Titan \\
 &  &  & & 2016 06 12 & J1517-2422 & J1551-1755 & J1517-2422 \\
J16090075-1908526 & 48 & 15 -- 704 & 21 & 2016 05 17 & J1517-2422 & J1551-1755 & Titan \\
 &  &  & & 2016 05 23 & J1517-2422 & J1551-1755 & Titan \\
 &  &  & & 2016 06 12 & J1517-2422 & J1551-1755 & J1517-2422 \\
J16142029-1906481 & 48 & 15 -- 704 & 21 & 2016 05 17 & J1517-2422 & J1551-1755 & Titan \\
 &  &  & & 2016 05 23 & J1517-2422 & J1551-1755 & Titan \\
 &  &  & & 2016 06 12 & J1517-2422 & J1551-1755 & J1517-2422 \\
J16123916-1859284 & 48 & 15 -- 704 & 21 & 2016 05 17 & J1517-2422 & J1551-1755 & Titan \\
 &  &  & & 2016 05 23 & J1517-2422 & J1551-1755 & Titan \\
 &  &  & & 2016 06 12 & J1517-2422 & J1551-1755 & J1517-2422 \\
IM Lup & 41 & 15 -- 630 &12  & 2016 05 01 & J1517-2422 & J1610-3958 & Titan \\
AS 209 & 41 & 15 -- 640 & 11 & 2016 05 15 & J1517-2422 & J1629-1720 & J1517-2422 \\
HD 163296 &   50 & 15 -- 705 & 60 & 2016 05 17 & J1924-2914 & J1743-1658 & Pallas \\
   & & & & 2016 06 21 & J1924-2914 & J1743-1658 & J1924-2914, J1733-1304 \\
\hline
\multicolumn{8}{c}{Cycle 4 -- 2016.1.00627.S} \\
\hline
MWC 480 & 44 & 15 -- 704 & 22 & 2016 12 01 & J0510+1800 & J0512+2927 & J0510+1800 \\
LkCa 15 & 47 & 15 -- 704 & 24 & 2016 12 01 & J0510+1800 & J0426+2327 & J0510+1800 \\
 & & & & 2016 12 02 & J0510+1800 & J0426+2327 & J0510+1800 \\
  & & & & 2016 12 03 & J0237+2848 & J0426+2327 & J0423-0120 \\
DO Tau & 47 & 15 -- 704 & 26 & 2016 12 01 & J0510+1800 & J0426+2327 & J0510+1800 \\
 & & & & 2016 12 02 & J0510+1800 & J0426+2327 & J0510+1800 \\
   & & & & 2016 12 03 & J0237+2848 & J0426+2327 & J0423-0120 \\
DM Tau & 47 & 15 -- 704 & 22 & 2016 12 01 & J0510+1800 & J0426+2327 & J0510+1800 \\
 & & & & 2016 12 02 & J0510+1800 & J0426+2327 & J0510+1800 \\
   & & & & 2016 12 03 & J0237+2848 & J0426+2327 & J0423-0120 \\
CI Tau & 47 & 15 -- 704 & 24 &  2016 12 01 & J0510+1800 & J0426+2327 & J0510+1800 \\
 & & & & 2016 12 02 & J0510+1800 & J0426+2327 & J0510+1800 \\
   & & & & 2016 12 03 & J0237+2848 & J0426+2327 & J0423-0120 \\
\hline
\multicolumn{8}{c}{Cycle 3 -- 2015.00671.S} \\
\hline
V4046 Sgr & 38 & 15 -- 705  & 9 & 2016 06 20 & J1924-2914 & J1816-3052 & J1733-1304 \\
\hline
\multicolumn{8}{c}{Cycle 2 -- 2013.1.00226.S} \\
\hline
V4046 Sgr & 36 & 21 -- 558 & 21 & 2015 05 13 & J1924-2914 & J1826-2924 & Titan 
\enddata
\end{deluxetable}

\begin{deluxetable}{lcccccc} 
	\tabletypesize{\footnotesize}
	\tablecaption{HCN SMA observations \label{tab:sma_details}}
	\tablecolumns{7} 
	\tablewidth{\textwidth} 
	\tablehead{
		\colhead{Date}           &
		\colhead{Sources}        &
		\colhead{\# Ant.}              & 
		\colhead{Baselines (m)}  &
		\colhead{Bandpass cal.}              &
		\colhead{Flux cal.}                       &
		\colhead{Gain cal.}                       
		}
\startdata
\multicolumn{7}{c}{2017B-S036} \\
\hline
2017 11 04 & DO Tau & 8 & 44 -- 226 &  3c84, 3c454.3 & Uranus & 0510+180, 3c111\\
2017 11 05 & CI Tau & 8 & 44 -- 226 &   3c84, 3c454.3 & Uranus, Mars & 0510+180, 3c111\\
2017 11 11 & DM Tau & 8 & 44 -- 226 & 3c84, 3c454.3 & Uranus & 0510+180, 0449+113 \\
2017 12 11 & CI Tau, DO Tau, DM Tau, & 8 & 16 -- 77 & 3c279, 3c84 & Uranus, Callisto & 0510+180, 3c11 \\
 & MWC 480, LkCa 15 \\
2018 01 01 & CI Tau, DO Tau, DM Tau, & 8 & 16 -- 77 & 3c279, 3c84 & Uranus, Neptune & 0510+189, 3c111 \\
 & MWC 480, LkCa 15 \\
2018 01 01 & CI Tau, DO Tau, DM Tau, & 8 & 16 -- 77 & 3c279, 3c84 & Uranus & 0510+180, 3c111\\
 & MWC 480, LkCa 15 \\
\enddata
\end{deluxetable}

\begin{deluxetable}{lccccc} 
	\tabletypesize{\footnotesize}
	\tablecaption{H$^{13}$CN, C$^{18}$O, CO, and C$_2$H Line Observations (Detections and Nondetection Upper Limits) \label{tab:fluxes_other}}
	\tablecolumns{6} 
	\tablewidth{\textwidth} 
	\tablehead{
		\colhead{Source}               &
		\colhead{Beam}                 &
		\colhead{Beam PA}           &
		\colhead{Channel rms$^a$}      &
		\colhead{Mom. Zero rms$^b$}  &
		\colhead{Int. Flux Density$^c$}  \\ 
		\colhead{}                           & 
		\colhead{('')}                        &
		\colhead{($^{\rm{o}}$)}          &
		\colhead{(mJy beam$^{-1}$)} &
		\colhead{(mJy beam$^{-1}$ km s$^{-1}$)} & 
		\colhead{(mJy km s$^{-1}$)}                      
		}
\startdata
\multicolumn{6}{c}{H$^{13}$CN J = 3--2} \\
\hline
AS 209 & 0.49 $\times$ 0.48 & -10.6 & 4.2 & 5.7 & 174 $\pm$ 33\\
CI Tau & 0.60 $\times$ 0.49 & -22.2 & 2.8 & 4.5 & $<$ 71\\
DM Tau & 0.57 $\times$ 0.49 & -31.6 & 2.9 & 4.3 & $<$ 80\\
DO Tau & 0.64 $\times$ 0.49 & -23.0 & 2.7 & 4.3 & $<$ 67\\
HD 163296 & 0.57 $\times$ 0.44 & 87.8 & 2.6 & 4.5 & 113 $\pm$ 36\\
LkCa 15 & 0.60 $\times$ 0.49 & -25.3 & 2.7 & 4.3 & 186 $\pm$ 29\\
MWC 480 & 0.74 $\times$ 0.47 & 14.3 & 3.9 & 6.7 & 165 $\pm$ 36\\
V4046 Sgr & 0.58 $\times$ 0.47 & -86.7 & 3.4 & 6.4 & 975 $\pm$ 100\\
\hline
\multicolumn{6}{c}{C$^{18}$O J = 2--1} \\
\hline
AS 209 & 0.58 $\times$ 0.57 & 56.8 & 5.7 & 7.6 & 403 $\pm$ 49\\
CI Tau & 0.78 $\times$ 0.54 & -32.2 & 3.3 & 4.6 & 591 $\pm$ 63\\
DM Tau & 0.73 $\times$ 0.54 & -37.2 & 3.4 & 4.7 & 1116 $\pm$ 110\\
DO Tau & 0.84 $\times$ 0.54 & -31.1 & 3.2 & 4.1 & 257 $\pm$ 30\\
HD 163296 & 0.66 $\times$ 0.56 & -61.7 & 3.4 & 6.7 & 5910 $\pm$ 590\\
IM Lup & 0.64 $\times$ 0.47 & 72.1 & 3.1 & 4.5 & 1225 $\pm$ 120\\
LkCa 15 & 0.61 $\times$ 0.47 & -12.7 & 2.9 & 4.0 & 589 $\pm$ 63\\
MWC 480 & 0.72 $\times$ 0.47 & -13.5 & 2.8 & 4.6 & 1486 $\pm$ 150\\
HD 143006 & 0.77 $\times$ 0.52 & 78.4 & 3.3 & 3.9 & 144 $\pm$ 19\\
J1604-2130 & 0.77 $\times$ 0.51 & 78.4 & 3.2 & 3.3 & 1377 $\pm$ 140\\
V4046 Sgr & 0.82 $\times$ 0.53 & 85.5 & 2.8 & 5.7 & 1203 $\pm$ 120\\
\hline
\multicolumn{6}{c}{$^{12}$CO J = 2--1} \\
\hline
J1609-1908 & 0.72 $\times$ 0.47 & 73.3 & 3.9 & 8.0 & 841 $\pm$ 88\\
J1612-1859 & 0.71 $\times$ 0.47 & 73.4 & 3.7 & 9.0 & 1586 $\pm$ 170\\
J1614-1906 & 0.71 $\times$ 0.47 & 74.2 & 4.0 & 9.0 & 1444 $\pm$ 150\\
\hline
\multicolumn{6}{c}{C$_2$H N = 3--2, J = $\frac{5}{2}$ -- $\frac{3}{2}$, F = 3 --2 and 2--1} \\
\hline
AS 209 & 0.65 $\times$ 0.55 & 77.7 & 10.2 & 19.1 & 2325 $\pm$ 250\\
CI Tau & 0.59 $\times$ 0.49 & -23.0 & 3.8 & 6.3 & 743 $\pm$ 82\\
DM Tau & 0.57 $\times$ 0.49 & -32.4 & 4.0 & 6.2 & 1453 $\pm$ 150\\
DO Tau & 0.63 $\times$ 0.49 & -23.3 & 3.7 & 6.6 & $<$ 73\\
HD 163296 & 0.55 $\times$ 0.47 & 76.8 & 2.5 & 5.8 & 3320 $\pm$ 340\\
IM Lup & 0.56 $\times$ 0.55 & 2.0 & 8.0 & 13.0 & 988 $\pm$ 130\\
LkCa 15 & 0.59 $\times$ 0.49 & -26.3 & 3.9 & 6.5 & 1675 $\pm$ 170\\
MWC 480 & 0.73 $\times$ 0.47 & 14.6 & 5.5 & 10.5 & 1569 $\pm$ 160\\
HD 143006 & 0.58 $\times$ 0.46 & 87.8 & 5.7 & 9.0 & 277 $\pm$ 47\\
J1604-2130 & 0.56 $\times$ 0.45 & -87.9 & 5.8 & 8.2 & 1926 $\pm$ 200\\
J1609-1908 & 0.56 $\times$ 0.45 & -86.5 & 5.7 & 11.1 & 591 $\pm$ 74\\
J1612-1859 & 0.56 $\times$ 0.45 & -87.0 & 5.6 & 11.6 & $<$ 90\\
J1614-1906 & 0.56 $\times$ 0.45 & -86.8 & 5.6 & 11.7 & $<$ 97\\
V4046 Sgr & 0.97 $\times$ 0.70 & 83.2 & 3.9 & 6.9 & 2206 $\pm$ 230\\
\enddata
\tablenotetext{a}{For 0.5 km s$^{-1}$ channel widths.  $^b$Median rms; see Section \ref{sec:datared}.  $^c$3$\sigma$ upper limits are reported for nondetections.  Uncertainties are derived by bootstrapping, added in quadrature with a 10\% calibration uncertainty. }
\end{deluxetable}

\clearpage
\FloatBarrier

\section{Hyperfine fitting: example spectra}
\label{sec:hf_ex}
\begin{figure}
\begin{centering}
	\includegraphics[width=0.9\linewidth]{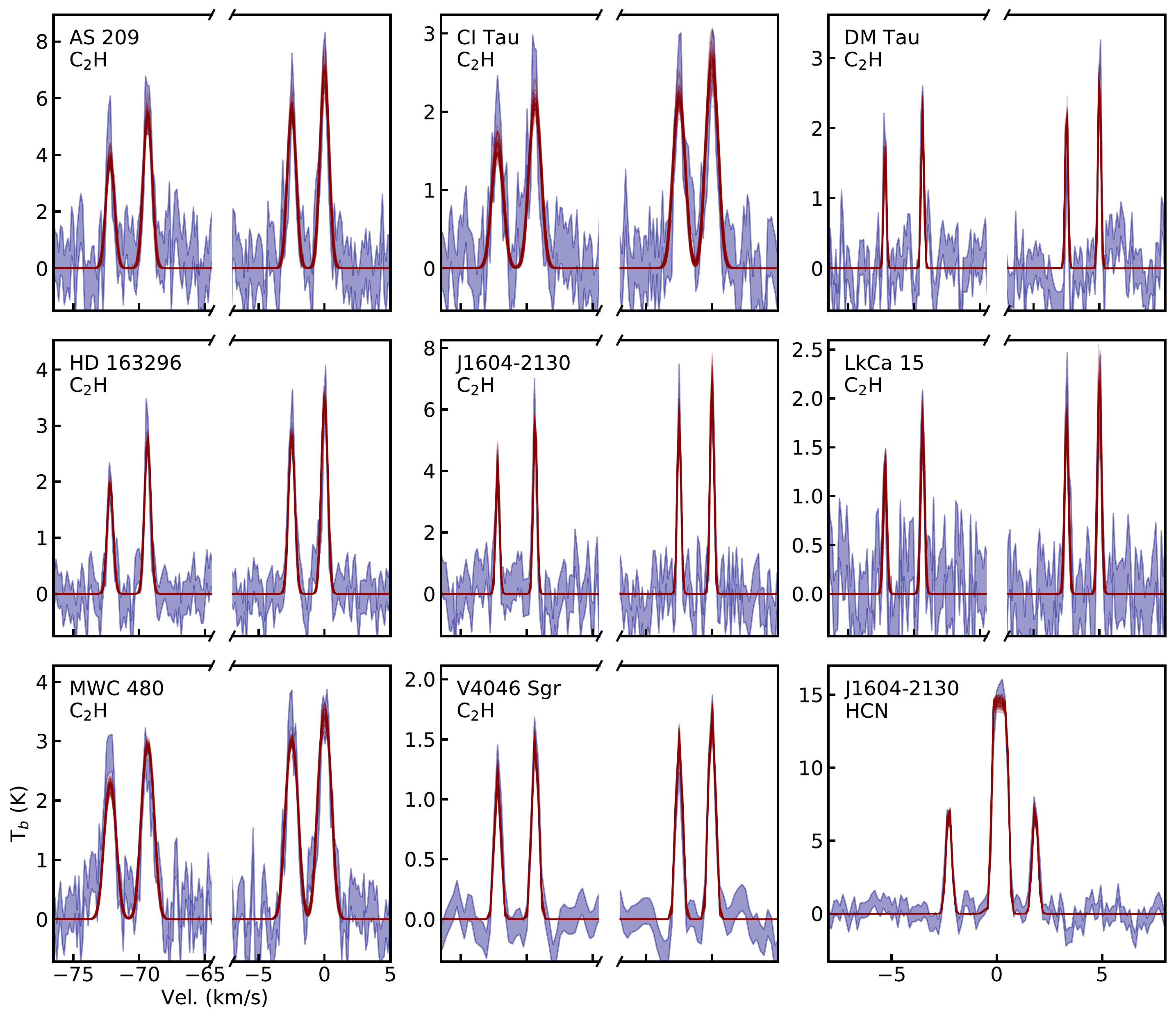}
	\caption{Example C$_2$H and HCN hyperfine fits in the sources with sufficient SNR and spectral resolution to perform fitting.  Blue shaded regions show the uncertainty range of the observational data, and red lines show draws from the fit posteriors.}
\end{centering}
\label{fig:ex_hf_fits}
\end{figure}

Figure \ref{fig:ex_hf_fits} shows example spectra fits using the hyperfine method described in Section \ref{sec:hf}.  The C$_2$H and HCN spectral line data can be found in Table \ref{tab:linedat}.  

\FloatBarrier

\end{document}